\documentclass[3p,12pt]{elsarticle}
\usepackage{lineno}
\modulolinenumbers[5]
\usepackage{graphicx}
\usepackage{amssymb}
\usepackage{bm,amsmath}
\usepackage{caption}
\usepackage{subeqnarray}
\usepackage{multirow}
\usepackage{lscape}
\usepackage{rotating}
\usepackage{subcaption}
\usepackage{lineno}

\usepackage{color,ulem}

\journal{Journal of computational physics}

\begin{document}

\title{ Accurate and efficient computation of the Boltzmann equation for Kramer's problem} 
	\author{Wei Su\fnref{label1,f1}}%
	\author{Peng Wang\fnref{label1,f1}}%
	\author{Haihu Liu \fnref{label2}}%
	\author{Lei Wu\corref{cor1}\fnref{label1}}%
	\ead{lei.wu.100@strath.ac.uk}
	\address[label1]{James Weir Fluids Laboratory, Department of Mechanical and Aerospace Engineering, University of Strathclyde, Glasgow G1 1XJ, UK}
	\address[label2]{School of Energy and Power Engineering, Xi'an Jiaotong University, 28 West Xianning Road, Xi'an 710049, China}
	\fntext[f1]{Wei Su and Peng Wang contribute equally}
		\cortext[cor1]{Corresponding author}
	\date{\today}


\begin{abstract}

The Kramer's problem is one of the fundamental problems of rarefied gas dynamics, which has been investigated extensively based on the linearized Boltzmann equation (LBE) of hard-sphere molecules and simplified kinetic model equations. However, how the different intermolecular potentials affect the viscous slip coefficient and the structure of Knudsen layer remains unclear. Here, a novel synthetic iteration scheme (SIS) is developed for the LBE to find solutions to Kramer's problem accurately and efficiently: the velocity distribution function is first solved by the conventional iterative scheme, then it is modified such that in each iteration i) the flow velocity is guided by an ordinary differential equation that is asymptotic-preserving at the Navier-Stokes limit and ii) the shear stress is equal to the average shear stress. Based on the Bhatnagar-Gross-Krook model, the SIS is assessed to be efficient and accurate.
Then we investigate the Kramer's problem for gases interacting through the inverse power-law, shielded Coulomb, and Lennard-Jones potentials, subject to diffuse-specular and Cercignani-Lampis gas-surface boundary conditions. When the tangential momentum accommodation coefficient (TMAC) is not larger than one, the Knudsen layer function is strongly affected by the potential, where its value and width increase with the effective viscosity index of gas molecules. Moreover, the Knudsen layer function exhibits similarities among different values of TMAC when the intermolecular potential is fixed. For Cercignani-Lampis boundary condition with TMAC larger than one,  both the viscous slip coefficient and Knudsen layer function are affected by the intermolecular potential, especially when the ``backward'' scattering limit is approached. With the asymptotic theory by Jiang and Luo (\textit{J. Comput. Phys}., vol. 316, 2016, pp. 416--434) for the singular behavior of the velocity gradient in the vicinity of the solid surface, we find that the whole Knudsen layer function can be well fitted by the power series $\sum_{n=0}^{2}\sum_{m=0}^{2}c_{n,m}x^n(x\ln x)^m$, where $x$ is the distance to the solid surface.  Finally, the experimental data of the Knudsen layer profile are explained by the LBE solution with proper values of the viscosity index and TMAC.
\end{abstract}


	\maketitle

\section{Introduction}

\begin{figure}
	\vskip 0.3cm
	\centering
	\includegraphics[width=0.5\textwidth]{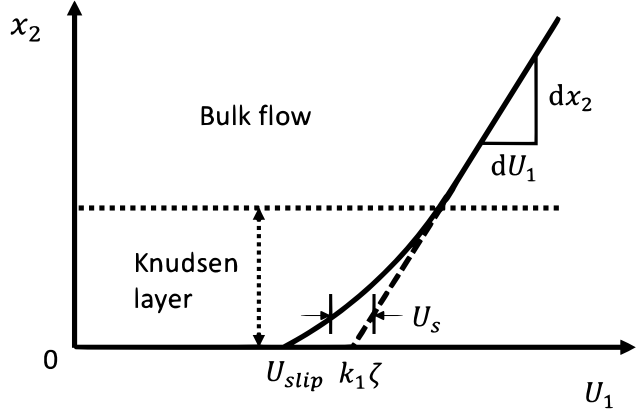}
	\caption{Schematic diagram of the Knudsen layer in the Kramer's problem. The velocity defect (Knudsen layer function) $U_s$ describes the deviation of the linearly extrapolated velocity (dash line) in the bulk region from the true velocity (solid line). The velocity slope in the bulk region is  $(\text{d}U_1/\text{d}x_2)|_{x_2\rightarrow\infty}=:k_1$, the slip length is $\zeta$, while the viscous slip coefficient is defined as $\bar{\zeta}=\zeta/\lambda_e$, where $\lambda_e$ is the equivalent mean free path of gas molecules.  }
	\label{fig:kramers_dia}
\end{figure}

The Kramer's problem is fundamental to solutions of almost all momentum transfer problems of rarefied gas dynamics~\cite{kramers1949behaviour}. As illustrated in Fig.~\ref{fig:kramers_dia}, when the planar wall at $x_2=0$ moves slowly in the horizontal direction, a nonlinear velocity profile and a finite slip velocity develop near the surface. This kinetic boundary layer, as known as the Knudsen layer, has a thickness of several mean free path $\lambda$ of gas molecules. Due to the infrequent gas-gas interactions, the flow is essentially rarefied, so that the conventional Navier-Stokes equations only work in the bulk region but break down in the Knudsen layer. The linearized Boltzmann equation (LBE) can be used to study this problem. For engineering applications, however, Navier-Stokes equations are still preferred when the Knudsen number $Kn$ (the ratio of $\lambda$ to the dimension of flow domain) is small, due to its distinct computational advantage over the LBE in six-dimensional phase space.

Many efforts have been made to predict the rarefied gas flow, through incorporating the rarefaction effects caused by the presence of solid surface into hydrodynamic equations~\cite{lockerby2008modelling}. For isothermal flows at small $Kn$, it is adequate to apply the velocity slip boundary condition (BC) to  Navier-Stokes equations. In this case, the viscous slip coefficient ($\bar{\zeta}$, VSC), as defined in Fig.~\ref{fig:kramers_dia}, is needed. The first estimation $\bar{\zeta}(\alpha)=(2-\alpha)/{\alpha}$ is proposed by Maxwell using insightful physical arguments~\cite{maxwell1879vii}. Here, $\alpha$ is the tangential momentum accommodation coefficient (TMAC) describing the fraction of diffusely reflected molecules at the solid surface, while the rest of molecules are reflected specularly. Almost one hundred years later, using a variational approach for the LBE and diffuse-specular gas-surface BC, Loyalka obtained the VSC which is generalized into the following form~\cite{loyalka1968momentum,sharipov1998data}:
	\begin{equation}\label{eq:loyalka}
	\bar{\zeta}(\alpha)=\frac{2-\alpha}{\alpha}\left[\bar{\zeta}(1)-0.1211(1-\alpha)\right].
	\end{equation}
Subsequently, lots of investigations were performed to calculate the VSC by numerically solving the LBE and its simplified model equations. It is found that VSCs from the LBE and its kinetic model equations have a relative difference less than $3\%$ when the effective TMAC is fixed~\cite{sharipov1998data}. It should be noted that, although the influence of intermolecular potentials on VSC has been assessed by Loyalka using the variational results~\cite{loyalka1990slip} and by Sharipov comparing the results from different literatures~\cite{sharipov2011data}, a systematic investigation of the role of the intermolecular potential on the VSC under different gas-surface interactions on the basis of the highly accurate Boltzmann solutions is still absent.

When $Kn$ becomes appreciable, Navier-Stokes equations may be still used, but in addition to the velocity slip BC the viscosity is modified to be a function of the distance to solid surfaces. In this case, the structure of the Knudsen layer provides a critical information to formulate the effective viscosity. Lockerby \textit{et al.} first proposed a curve-fitted approximation to the Knudsen layer function (KLF) as~\cite{lockerby2005capturing}
	\begin{equation}\label{eq:loc_vis}
	U_s\left(x\right)\approx\frac{7}{20(1+x)^2},
	\end{equation}
where $x$ is the distance to the solid surface normalized by the mean free path $\lambda$. Although the KLF is fitted from a temperature jump problem instead of the shear problem, it is found that the Navier-Stokes equations with the effective viscosity can predict the velocity profiles in Poiseuille and Couette flows, up to $Kn=0.4$. Later, by fitting the data from the LBE solution of hard-sphere (HS) gas and the direct simulation Monte Carlo method for Couette flow~\cite{ohwada1989numerical}, Lilley~$\&$~Sader obtained a power-law KLF~\cite{lilley2007velocity,lilley2008velocity}:
	\begin{equation}
	U_s\left(x\right)=U_s(0)-Cx^n,
	\end{equation}
where $C$ is a constant and the exponent $n\approx0.82$. Although the fitting is carried out in the region $0.1{\lesssim}x\lesssim1$, they predicted the power-law divergence of the velocity gradient in the vicinity of the solid surface, that is, $dU_s/dx\rightarrow\infty$ as $x\rightarrow0$.


The singular behavior of the velocity gradient at the planar surface is rigorously proved by Takata \& Funagane~\cite{takata2013singular}, when analyzing the thermal transpiration based on the LBE of HS molecules. However, instead of the power-law divergence, they found the logarithmic divergence of the velocity gradient; that is, the spatial singularity is not stronger than $\ln{}x$ in the vicinity of the solid surface. This conclusion is confirmed by Jiang \& Luo who, through the asymptotic analysis of the Bhatnagar-Gross-Krook (BGK) model~\cite{bhatnagar1954model}, found that the velocity profile of Couette flow near the solid surface can be described by the following power series~\cite{Jiang2016JCP}
	\begin{equation}\label{eq:vel_fit}
	U_s(x)=\sum_{n=0}^{N}\sum_{m=0}^{M}c_{n,m}x^n(x\ln{x})^m, \quad{} x\rightarrow0. 
	\end{equation}

It should be noted that most contributions to the Kramer's problem focused mainly on the diffuse-specular BC and HS molecules (or simplified kinetic models). How intermolecular potentials (such as the inverse power-law, shielded Coulomb, and Lennard-Jones potentials) and other gas-kinetic BCs affect the VSC and KLF remains unclear. This paper is dedicated to addressing these questions through the numerical simulation of the LBE. We emphasis that, however, the numerical method to finding the KLF at small values of $Kn$ is not easy. For instance, the results provided by Takata \& Funagane are limited to $Kn\gtrsim0.6$, since the computational cost to find the steady-state solution of the kinetic equations becomes extremely large for small Knudsen numbers~\cite{takata2013singular}; however, this relative large value of $Kn$ is unfortunately not small enough to avoid the interference between Knudsen layers. In the present paper, we first develop an efficient and accurate method to solve the LBE, and then investigate the role of intermolecular potentials and gas-surface BCs on the VSC and KLF.

The remainder of the paper is organized as follows. In \S~\ref{sectionII}, the LBE for the steady Couette flow of a monatomic gas and various kinetic BCs are introduced. In \S~\ref{secIII}, a synthetic iteration scheme is developed to boost the convergence in finding the steady-state solution of the Couette flow in the near-continuum regime. In \S~\ref{sec:results}, influences of  intermolecular potentials and gas-kinetic BCs on the VSC and KLF as well as the singularity of the velocity gradient near the solid surface and the similarity of the KLF are investigated. In \S~\ref{sec:exp}, experimental results given by Reynolds \textit{et al.}~\cite{reynolds1974velocity} are properly explained. The paper closes with some finial comments in \S~\ref{sec:summary}.

\section{The linearized Boltzmann equation }\label{sectionII}

Consider the steady Couette flow of a monatomic gas between two infinite parallel plates located at $x_2=0$ and $x_2=1$. The top plate moves along the $x_1$ direction with the velocity $V_w$, while the bottom plate moves with the opposite velocity. Both plates are maintained at a fixed temperature $T_w$. This Couette flow can be used to study the Kramer's problem, provided that the distance between the two plates is large enough so that there is no interference between the Knudsen layers near the two plates~\cite{sharipov2011data}.

When $V_w$ is far smaller than the most probable speed (${v_m}=\sqrt{2k_BT_w/m}$, where $k_B$ is the Boltzmann constant and $m$ is the gas molecular mass) of the gas molecules, the velocity distribution function of gas molecules can be linearized around the global equilibrium distribution function $f_{eq}(\textbf{v})={\pi^{-3/2}}{\exp(-|\textbf{v}|^2)}$ as:
\begin{equation}
f(x_2,\textbf{v})=f_{eq}(\textbf{v})+\frac{V_w}{v_m}h(x_2,\textbf{v}), 
\end{equation}
where $\textbf{v}=(v_1,v_2,v_3)$ is the molecular velocity and $h(x_2,\textbf{v})V_w/v_m$ is the small perturbance ($h$ is not necessary smaller compared to $f_{eq}$). The LBE for $h(x_2,\textbf{v})$ is:
\begin{equation}\label{LBE}
v_2\frac{\partial
	{h}}{\partial{x_2}}=L(h,f_{eq}),
\end{equation}
where the linearized Boltzmann collision operator is~\cite{lei_Jfm}:
\begin{equation}\label{LBE_collision}
L=\underbrace{\iint B(\theta,|\textbf{u}|)  [f_{eq}(\textbf{v}')h({\textbf{v}}'_{\ast})+f_{eq}(\textbf{v}'_\ast)h({\textbf{v}}')-f_{eq}(\textbf{v})h({\textbf{v}}_\ast)]d\Omega d{\textbf{v}}_\ast}_{L^+}-\nu_{eq}(\textbf{v})h(\textbf{v}),
\end{equation}
and the equilibrium collision frequency is
\begin{equation}
\nu_{eq}(\textbf{v})=\iint B(|\textbf{v}-\textbf{v}_\ast|,\theta) f_{eq}(\textbf{v}_\ast)d\Omega{}d{\textbf{v}}_\ast.
\end{equation}

Note that in the above LBE, the coordinate $x_2$ has been normalized by the distance between the two plates $H$, the molecular velocity $\textbf{v}$ has been normalized by the most probable speed $v_m$, and velocity distribution functions $f_{eq}$ and $h$ have been normalized by $n_0/v_m^3$, where $n_0$ is the average number density of the gas molecules between the parallel plates. The relative velocity of the two molecules before binary collision is $\textbf{u}=\textbf{v}-\textbf{v}_\ast$, and $\Omega$ is a unit vector along the relative post-collision velocity $\textbf{v}'-\textbf{v}'_\ast$. The deflection angle $\theta$ between the pre- and post-collision relative velocities satisfies $\cos\theta=\Omega\cdot{\textbf{u}}/|\textbf{u}|$, $0\le\theta\le\pi$. Finally, $B(\theta,|\textbf{u}|)=|\textbf{u}|\sigma$ is the collision kernel, with $\sigma$ being the differential cross-section that is determined by the intermolecular potential.

In the present paper, we consider the inverse power-law potentials, where the collision kernels are modeled as~\cite{Lei2013,lei_Jfm}
\begin{equation}\label{collision_kernel}
B(|{\textbf{u}}|,\theta)=\frac{|{\textbf{u}}|^{2(1-\omega)}}{K}	{\sin^{\frac{1}{2}-\omega}\left(\frac{\theta}{2}\right)\cos^{\frac{1}{2}-\omega}\left(\frac{\theta}{2}\right)},
\end{equation}
with $\omega$ being the viscosity index (i.e. the shear viscosity $\mu$ of the gas is proportional to $T^\omega$) and $K$ some normalization constants~\cite{lei_Jfm}. HS and Maxwell molecules have $\omega=0.5$ and 1, respectively. Note that this type of collision kernel cannot describe the charged molecules interacting through the Coulomb potential with $\omega=2.5$~\cite{Lei2013}. As discussed in the Chapter 10 of Ref.~\cite{CE}, in reality, however, charged molecules interact through the shielded Coulomb potential:
\begin{equation}
U'(\rho')=\epsilon\frac{\lambda_d}{\rho'}\exp\left(-\frac{\rho'}{\lambda_d}\right),
\end{equation}
where $\epsilon$ is related to the strength of the potential, $\rho'$ is the intermolecular distance, and $\lambda_d$ is the Debye shielding length. For simplicity, we only consider the single-species charged molecules interacting through the repulsive force. We also consider noble gases interacting through the following Lennard-Jones potentials:
\begin{equation}
U'(\rho')=4\epsilon\left[\left(\frac{d}{\rho'}\right)^{12}-\left(\frac{d}{\rho'}\right)^6\right],
\end{equation}
where $d$ is the distance at which the potential is zero. The details of implementation of the Lennard-Jones potentials in FSM can be found in Ref.~\cite{Wu:2015yu}. 

The differential cross-section for the above shielded Coulomb and Lennard-Jones potentials can be calculated according to Sharipov \& Bertoldo~\cite{Sharipov2009a}. Then the linearized Boltzmann collision operator Eq.~\eqref{LBE_collision} can be solved by the fast spectral method developed by the authors~\cite{Wu:2015yu}.

To fully determine the gas dynamics in spatially-inhomogeneous problems, the gas-surface BC is needed. The general form of the BC, which specifies the relation between the velocity distribution function $f(\textbf{v})$ of the reflected and incident gas molecules at the solid surface, is given below:
\begin{equation}\label{label_general}
v_nf(\textbf{v})=\int_{v_n'<0} |v_n'| R(\textbf{v}' \rightarrow \textbf{v}) f(\textbf{v}') d\textbf{v}', \quad  v_n>0,
\end{equation}
where $\textbf{v}'$ and $\textbf{v}$ are velocities of the incident and reflected molecules, respectively, $v_n$ is the normal component of the molecular velocity $\textbf{v}$ directed into the gas, and $R(\textbf{v}' \rightarrow \textbf{v})$ is the non-negative scattering kernel.

The most popular gas-surface BC is the diffuse-specular one, with the scattering kernel reading:
	\begin{equation}
	R_M(\textbf{v}' \rightarrow \textbf{v})=\alpha_M \frac{m^2 {v}_n}{2\pi(kT_w)^2}\exp\left(-\frac{m  \textbf{v}^2 }{2kT_w}\right)+\left(1-\alpha_M\right)\delta\left( \textbf{v}'-\textbf{v} +2\textbf{n} {v}_n \right),
	\end{equation}
where the constant $\alpha_M$ is the TMAC, with a value in the range of $0\leq \alpha_M \leq 1$, and $\delta$ is the Dirac delta function.  Purely diffuse reflection has $\alpha_M = 1$. The BC proposed by Cercignani \& Lampis~\cite{cercignani1971kinetic} has also been widely used, which reads:
	\begin{equation}\begin{aligned}
	R_{CL}\left(\textbf{v}'\rightarrow \textbf{v} \right )= & \frac{m^2 v_n}{2\pi \alpha_n\alpha_t\left(2-\alpha_t\right)\left(kT_w\right)^2} I_0\left(\frac{\sqrt{1-\alpha_n}mv_nv_n'}{\alpha_nkT_w}\right) \\
	&\times \exp\left\{ -\frac{m\left[v_n^2+(1-\alpha_n)v_n'\right]^2}{2kT_w\alpha_n}-\frac{m\left[\textbf{v}_t-(1-\alpha_t)\textbf{v}_t'\right]^2}{2kT_w\alpha_t(2-\alpha_t)}\right\},
		\end{aligned}
	\end{equation}
where $\textbf{v}_t$ is the tangential velocity,  $I_0(x)=\int_{0}^{2\pi}\exp\left(x\cos\phi\right)\text{d}\phi/2\pi$, and $\alpha_n\in[0,1]$ and $\alpha_t\in[0,2]$ are the energy and momentum accommodation coefficients, respectively. When $\alpha_n=\alpha_t=1$ or $\alpha_n=\alpha_t=0$, the fully diffuse or specular BCs are recovered, respectively, while for $\alpha_n=0$ and $\alpha_t=2$, the Cercignani-Lampis scattering kernel descries ``backward'' scattering. Other types of BCs have also been proposed and discussed~\cite{WuStruchtrupJFM2017}, but for the Kramer's problem, as will be shown below, the two BCs are adequate to explain the experimental data of Reynolds \textit{et al.}~\cite{reynolds1974velocity}.


The macroscopic quantities of interest are the flow velocity normalized by $V_w$ and the shear stress normalized by $n_0k_BT_wV_w/v_m$, which can be calculated as
\begin{equation}
\begin{aligned}[b]
U_1=&\int{v_1h}d\textbf{v}, \\ P_{12}=&\int{2v_1v_2h}d\textbf{v}.
\end{aligned}
\end{equation}

\section{Numerical method: the synthetic iteration scheme}\label{secIII}

To resolve the singular behavior of the velocity gradient that occurs in the vicinity of the solid surface~\cite{takata2013singular,Jiang2016JCP}, high spatial resolution is required. This means that it is better to solve kinetic equations by time-implicit deterministic numerical method, otherwise the restriction on the Courant-Friedrichs-Lewy condition will render the time step extremely small and hence the computational cost enormous; also, the direct simulation Monte Carlo method will be expensive to resolve the velocity profile in very small cells near the solid surface. 

To avoid the interference of the two Knudsen layers, the mean free path of gas molecules should be sufficiently smaller than the distance between two parallel plates; or equivalently, the rarefaction parameter
\begin{equation}\label{delta}
\delta=\frac{H}{\lambda_e}, ~\lambda_e=\frac{\mu{(T_w)}v_m}{n_0k_BT_w},
\end{equation}
should be sufficiently large. Here the equivalent mean free path $\lambda_e =(2/\sqrt{\pi})\lambda$, where $\lambda$ is the mean free path of the gas molecules. Therefore, the Knudsen number is  $Kn=\sqrt{\pi}/2\delta$.

The integro-differential system Eq.~\eqref{LBE} is usually solved by the conventional iteration scheme. Given the value of $h^{(k)}(x_2,\textbf{v})$ at the $k$-th iteration step, the velocity distribution function at the next iteration step is calculated by solving the following equation~\cite{ohwada1989numerical,Lei2013,Wu:2015yu}:
\begin{equation}\label{LBE_iteration}
\nu_{eq}h^{(k+1)}+v_2\frac{\partial
	{h}^{(k+1)}}{\partial{x_2}}=L^+(h^{(k)},f_{eq}),
\end{equation}
where the derivative with respect to $x_2$ is usually approximated by a second-order upwind finite difference, and the collision operator in Eq.~\eqref{LBE_collision} can be calculated by the fast spectral method~\cite{lei_Jfm,Wu:2015yu} based on the velocity distribution function at the $k$-th iteration step. The process is repeated until relative differences between successive estimates of macroscopic quantities are less than a convergence criterion $\epsilon$.

The conventional iteration scheme is efficient for highly rarefied gas flows (when $\delta$ is very small), where converged solutions can be quickly found after several iterations. However, the number of iteration increases significantly when $Kn$ decreases (or $\delta$ increases), especially when the gas flow is in the near-continuum regime~\cite{Valougeorgis:2003zr,LeiJCP2017}. These behaviors are in fact a result of the competition between the molecular collision and streaming. In the free-molecular flow regime, gas molecules move in straight way (except the collision with solid surfaces) so that any disturbance at one point can be quickly felt by all other spatial points, so the exchange of ``information'' and hence the convergence is fast. However, for near-continuum flows, binary collisions dominate so that the exchange of information through streaming becomes very inefficient: the perturbance decays rapidly due to frequent binary collisions and takes a long time to be felt by other points. The implicit unified gas-kinetic scheme may be used to achieve fast convergence~\cite{zhuyajun2016}, however, currently there is no version developed for the LBE.

To have a convergence-accelerated scheme for the LBE, synthetic equations for the evolution of macroscopic flow variables that are asymptotic preserving the Navier-Stokes limit provides an alternative way to enhance the information exchange across the whole computational domain~\cite{LeiJCP2017}. For the Navier-Stokes equations which can be derived from the Boltzmann equation through the Chapman-Enskog expansion to the first-order of the Knudsen number, the governing equation for the flow velocity is
\begin{equation}\label{eq:vis}
\frac{ \partial {U_1} } {\partial{}x_2}=-\delta{P_{12}},
\end{equation}
where $P_{12}$ is a constant across the whole domain. For the LBE, the shear stress remains a constant (this can be easily proven by multiplying Eq.~\eqref{LBE} with $v_1$ and then integrating with respect to $\textbf{v}$), but the equation for $U_1$ contains high-order terms beyond the Navier-Stokes level. That is, the governing equation is in general can be expressed as
\begin{equation}\label{synthetic}
\frac{\partial{}U_1} {\partial{}x_2}=-\delta{P_{12}}+\text{High-order terms}.
\end{equation}

To obtain the synthetic equation Eq.~\eqref{synthetic} that will facilitate the fast convergence to the steady-state, we first rewrite Eq.~\eqref{LBE_collision} as $L=(L-L_{BGK})+L_{BGK}$, where
\begin{equation}
L_{BGK}=\delta[2U_1v_1f_{eq}-h]
\end{equation}
is the linearized collision operator of the BGK equation for Couette flow between two parallel plates~\cite{LeiJCP2017}. Then multiplying Eq.~\eqref{LBE} by $2v_1v_2$ and integrating the resulting equation with respect to the molecular velocity $\textbf{v}$, we obtain
\begin{eqnarray}\label{synthetic2}
\frac{\partial U_1}{\partial x_2}=-\delta{}P_{12}+\underbrace{\int{}2v_1v_2(L-L_{BGK})d\textbf{v}-\frac{\partial }{\partial x_2}\int(2v_2^2-1)v_1fd\textbf{v}}_{\text{High-order terms}}.
\end{eqnarray}

It is obvious that, in the near-continuum regime where $\delta\rightarrow\infty$, the high-order terms in the right-hand side of Eq.~\eqref{synthetic2} are negligible compared to $\delta{}P_{12}$, so that the derived synthetic equation is asymptotic preserving the Navier-Stokes limit. With this macroscopic equation to update the flow velocity, we devise the following new iteration scheme to find the steady-state solution of the LBE Eq.~\eqref{LBE} quickly:
\begin{itemize}

	\item Due to the symmetry condition $h(x_2,v_1,v_2,v_3)=-h(1-x_2,v_1,-v_2,v_3)$, the computational domain will be limited to $0\le{x_2}\le{1/2}$. When $h^{(k)}$ and $U_1^{(k)}$ are known at the $k$-th iteration, we calculate one of the high-order terms $H_1(x_2)=\int 2v_1v_2(L-L_{BGK})d\textbf{v}$. We also calculate the velocity distribution function $h^{(k+1/2)}$ according to the conventional iteration scheme Eq.~\eqref{LBE_iteration}, that is, we solve the following equation:
	\begin{equation}\label{syn_LBE}
	{\nu_{eq}}h^{(k+1/2)}+v_2\frac{\partial
		{h}^{(k+1/2)}}{\partial{x_2}}=L^+(h^{(k)},f_{eq}),
	\end{equation}
	by a second-order upwind finite difference in the bulk and a first-order upwind scheme at the solid surface~\cite{ohwada1989numerical}.

	\item From $h^{(k+1/2)}$, we calculate the flow velocity $U_1^{(k+1/2)}(x_2)$, the shear stress $P_{12}^{(k+1/2)}(x_2)$, and one of the high-order terms $H_2(x_2)=\int(2v_2^2-1)v_1fd\textbf{v}$. We also calculate the average shear stress as
	\begin{equation}\label{mean_stress}
	\bar{P}=2\int_0^{1/2}{}P_{12}^{(k+1/2)}dx_2.
	\end{equation}

	\item We obtain the flow velocity $U_1^{(k+1)}$ by solving Eq.~\eqref{synthetic2} with the symmetrical boundary condition $U_1(1/2)=0$, where $P_{12}$ is replaced by $\bar{P}$. That is,
	\begin{eqnarray}\label{synthetic3}
	\frac{\partial U_1^{(k+1)}(x_2)}{\partial x_2}=-\delta{}\bar{P}+H_1(x_2)-\frac{\partial{H_2(x_2)}}{x_2}.
	\end{eqnarray}

	\item The velocity distribution function $h(x_2,\textbf{v})$ is modified to incorporate the change of the macroscopic flow velocity. Meanwhile, the shear stress is adjusted to its mean value $\bar{P}$, for all spatial points. That is,
	\begin{equation}\label{guided}
	h^{(k+1)}=h^{(k+1/2)}+2\left(U_1^{(k+1)}-U_1^{(k+1/2)}\right)v_1f_{eq}+2\left(\bar{P}-P_{12}^{(k+1/2)}\right)v_1v_2f_{eq}.
	\end{equation}

	\item The above steps are repeated until convergence.\\
\end{itemize}

Since the gas kinetic equation is solved together with the macroscopic equation Eq.~\eqref{synthetic3} for flow velocity, the above scheme is called the synthetic iterative scheme (SIS). Note that although the SIS has been widely applied to the radiation transport processes~\cite{DSA2002} and rarefied gas flows driven by local pressure, temperature, and concentration gradients to overcome the slow convergence in the near-continuum flow regime~\cite{Valougeorgis:2003zr,Naris2005Pof,CircularSIS2013}, it is the first time that the SIS is developed for the linearized Couette flow.

\subsection{Numerical tests of efficiency and accuracy}\label{NumSingle}

\begin{figure}[t]
	\centering
	\includegraphics[scale=0.55,viewport=30 240 670 470,clip=true]{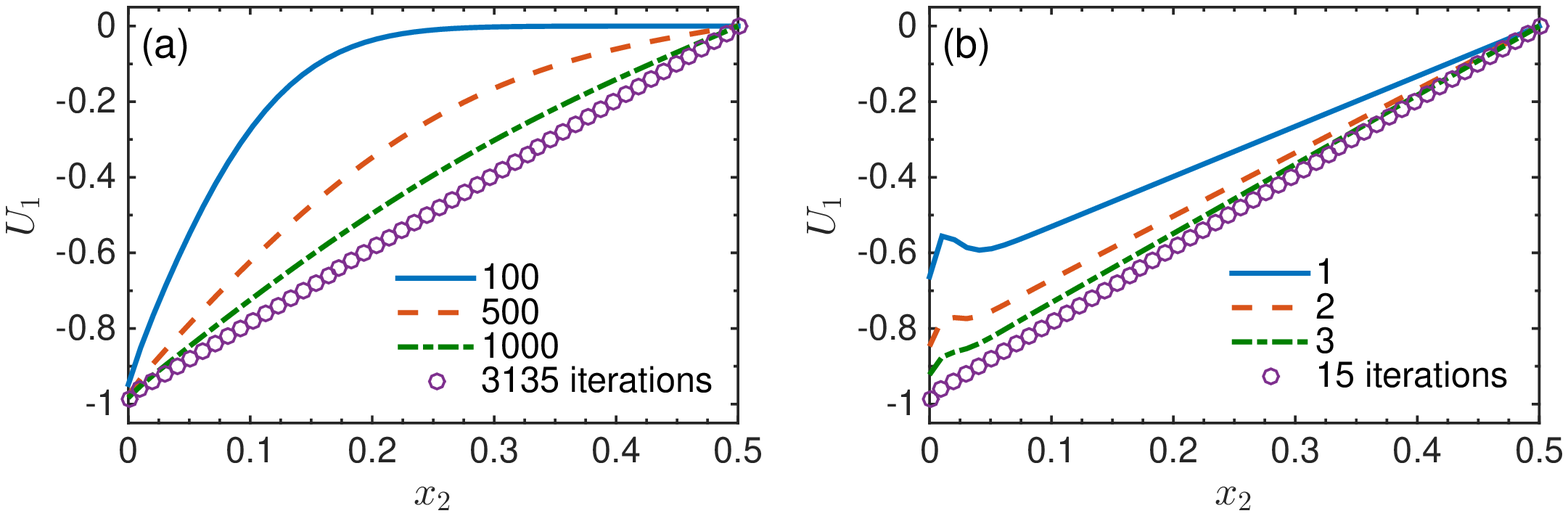}\\
	\includegraphics[scale=0.45,viewport=0 0 440 290,clip=true]{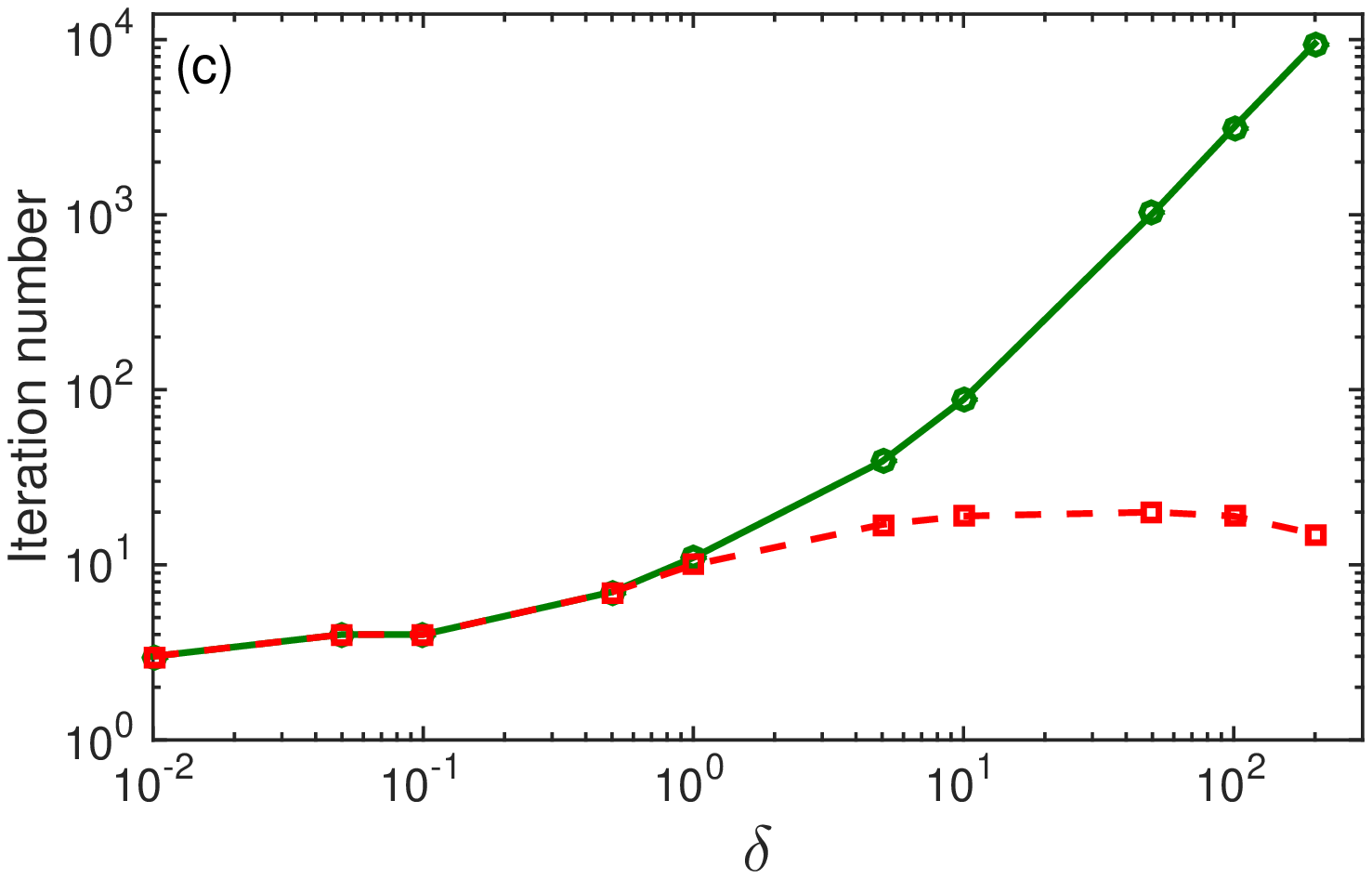}
	\caption{Profiles of the flow velocity at different iteration steps obtained from the conventional iteration scheme (a) and SIS (b), when $\delta=100$. Circles show the converged solution. (c) The total iteration number needed to obtain the converged solution as a function of the rarefaction parameter $\delta$, where circles and squares are the results from the conventional iteration scheme and SIS, respectively. The initial condition is $h(x_2,\textbf{v})=0$. The iteration is terminated when the maximum relative difference in the flow velocity between two consecutive iterations is less than $10^{-5}$.  }
	\label{Converg_profile}
\end{figure}

Numerical simulations are carried out to assess the efficiency and accuracy of the SIS. We consider the simple BGK kinetic model with the diffuse boundary condition as it has recently been solved with high accuracy~\cite{Jiang2016JCP,SaderPoF2012,li2015accurate}.

We first test the efficiency of the SIS. We choose the rarefaction parameter $\delta=100$ and discretize the half spatial space into 50 even-spaced points. The molecular velocity space $v_1$ and $v_3$ are discretized by the roots of the physicists' version of the fourth-order Hermite polynomial, while the molecular velocity $v_2$ is truncated to $[-6,6]$ and approximated by the non-uniform points~\cite{lei_Jfm,SuWeiPRE2017}:
\begin{equation}\label{nonuniform_v}
v_{2}=\frac{6}{(N_{v}-1)^\imath}(-N_{v}+1,-N_{v}+3,\cdots,{N_{v}-1})^\imath,
\end{equation}
which is useful to capture the discontinuity in the velocity distribution function near $v_2\sim0$. In this test we take $\imath=3$.

Figure~\ref{Converg_profile} compares the convergence history and speed (in terms of the number of iterative steps to reach the converged solution) of the SIS to the conventional iteration scheme. Starting from the initial guess $h(x_2,\textbf{v})=0$, the perturbance from the solid surface quickly adjusts the flow velocity near the solid surface towards the surface velocity in the conventional iteration scheme: from Fig.~\ref{Converg_profile}(a) we find that $U_1(0)$ is already very close to the final converged solution after 100 iterations. However, due to the frequent binary collision, such a perturbance slowly penetrates the bulk regime. Since the symmetry condition always guarantees $U_1(1/2)=0$, a large number of iterations are needed to alter the velocity profile between the solid surface and the center of the channel to be nearly linear. This situation is completely changed in the SIS, where the flow velocity is corrected to be nearly linear at each iteration according to the synthetic equation Eq.~\eqref{synthetic3}, which can be approximated by $\partial {U_1}/\partial{}x_2=-\delta\bar{P}$ when $\delta$ is large. Such a macroscopic governing equation allows the efficient exchange of information, and therefore fast convergence is realized in the whole computational domain, see Fig.~\ref{Converg_profile}(b).

As far as the convergence speed is concerned, we see from Fig.~\ref{Converg_profile}(c) that, when $\delta$ is small, i.e. in the free-molecular flow regime, the conventional iteration scheme and SIS are as efficient as each other, where the converged solutions are obtained within 5 iterations. As $\delta$ increases so that the flow enters the transition and near-continuum regimes, the iteration number of the conventional iteration scheme increases rapidly, while that of the SIS quickly reaches the saturation number of about 20. At $\delta=200$, the SIS is about 500 times more efficient than the conventional iteration scheme. The gain of using SIS becomes larger and larger as $\delta$ further increases.

\begin{table}
	\caption{Comparisons of the velocity at the solid surface $x_2=1$, the velocity derivative at the channel center $x_2=1/2$, and the shear stress between the results of Jiang \& Luo~\cite{Jiang2016JCP} and SIS. The linearized BGK equation is used. At each value of $\delta$, the data of Jiang \& Luo~\cite{Jiang2016JCP} and SIS are shown in the first and second rows, respectively. }
	\label{table_compare_luo}
	\centering

	\begin{tabular}{cccc}
\hline
		$1/\delta$	&  {$U_1(1)/2$}
		&  {$dU_1(1/2)/2dx_2$}  &    {$-P_{12}/4$}\\

\hline
		$0.003$
		& 0.497891535  & 0.993939801   & 1.490909702$\times10^{-3}$    \\
		& 0.497891548   & 0.993939827  & 1.490909741$\times10^{-3}$ \\ 

		$0.01$& 0.493069780   & 0.980081002   & 4.900405010$\times10^{-3}$   \\
		& 0.493069792   & 0.980081024   & 4.900405118$\times10^{-3}$   \\

		$0.1$ & 0.441224641   & 0.835285766   & 4.155607783$\times10^{-2}$   \\
		& 0.441224646   & 0.835285536   & 4.155607809$\times10^{-2}$   \\

		$1$   & 0.251861340   & 0.444228470   & 1.694625753$\times10^{-1}$   \\
		& 0.251861372   & 0.444228442   & 1.694625700$\times10^{-1}$  \\

		$10$  & 0.072922113   & 0.132195579   & 2.611624603$\times10^{-1}$   \\
		& 0.072922127   & 0.132195588   & 2.611624596$\times10^{-1}$ \\

		$100$ & 0.013430729   & 0.025200983   & 2.796682147$\times10^{-1}$   \\
		& 0.013430736   & 0.025200817   & 2.796682147$\times10^{-1}$\\
\hline
\end{tabular}\par

\end{table}

We then assess the accuracy of the SIS by comparing the solution of the integral equation derived from the linearized BGK  equation, which has the accuracy of at least 12 significant digits~\cite{Jiang2016JCP}. In order to capture the Knudsen layer near the solid surface, the spatial domain $0\le{}x_2\le1/2$ is divided into $N_s$ nonuniform sections, with most of the discrete points placed near the wall:
\begin{equation}\label{space_discrete}
x_2=(10-15s+6s^2)s^3,
\end{equation}
where $s=(0,1,\cdots,N_s)/2N_s$. The size of the smallest section is $1.2\times10^{-9}$ when $N_s=1000$. The iterations terminate when the maximum relative error in the flow velocity between two consecutive iterations
\begin{equation}
\epsilon=\text{max}\left|\frac{U_1^{(k+1)}(x_2)}{U_1^{(k)}(x_2)}-1\right|
\end{equation}
is less than $10^{-10}$; the point $U_1(1/2)$ is excluded since the velocity is always zero.

A comparison between the SIS and accurate results of~\cite{Jiang2016JCP} is tabulated in Table~\ref{table_compare_luo} for the linearized Couette flow. The molecular velocity $v_2$ is discretized according to Eq.~\eqref{nonuniform_v} with $N_v=64$ and $\imath=5$, while in the spatial discretization Eq.~\eqref{space_discrete} we choose $N_s=500$. Clearly our SIS has an accuracy of at least 6 significant digits. The accuracy can be further increased when more refined velocity and spatial grids are used.

\section{Numerical results of the linearized Boltzmann equation}\label{sec:results}

Using the accurate and efficient SIS, the LBE is solved for different intermolecular potentials, under different gas-surface BCs. In the numerical simulation, we set the rarefaction parameter to be $\delta=100$, so that the  distance between two plates is about 100 times as large as the mean free path of gas molecules; thus, the interference between the Knudsen layers near each plate is avoided. The molecular velocity $v_2$ is discretized according to Eq.~\eqref{nonuniform_v} with $N_v=128$ and $\imath=5$, while $v_1$ and $v_3$ are discretized by $32\times32$ uniform grids in the range of $[-6,6]$; in the spatial discretization we choose $N_s=500$ in Eq.~\eqref{space_discrete}. In the fast spectral approximation of the linearized Boltzmann collision operator Eq.~\eqref{LBE_collision}, the integral with respect to the solid angle $\Omega$ is calculated by the Gauss-Legendre quadrature with $M=8$, see equation~(39) in Ref.~\cite{Lei2013}. All these measures enable our results holding an accuracy of at least 6 significant digits.

When the steady-state solution is obtained, the velocity profile in the bulk region is linearly fitted by $U_{NS}=k_1x_2+k_0$ in the dimensionless form, where $k_0$ and $k_1$ are coefficients from the least square fitting. Then the KLF is calculated according to the following equation:
\begin{equation}\label{NS_fit}
U_s\left(\frac{x_2}{Kn}\right)=\frac{U_{NS}(x_2)-U_1(x_2)}{k_1Kn},
\end{equation}
and the VSC is calculated as
\begin{equation}\label{slip_coe}
\bar{\zeta}=-\frac{1+k_0}{\bar{P}}.
\end{equation}

In the numerical simulation, we find that $\delta=100$ is accurate enough to recover the KLF and VSC, when compared to the solution of $\delta=1000$. However, when $\delta=10$, that is, the  distance between two plates is roughly 10 times of the mean free path, two Knudsen layers interact with each other, which leads to an inaccurate KLF by using Eq.~\eqref{NS_fit}.

\subsection{ The viscous slip coefficient}\label{sec:coe}

Although a large number of VSCs have been computed from  kinetic model equations~\cite{sharipov2011data}, very few data are available based on the LBE for various intermolecular potentials, in particular of the highly accurate solutions. In this section, we study how the intermolecular potentials (including the inverse power-law, shielded Coulomb, and Lennard-Jones potentials) and gas-kinetic BCs (including the diffuse-specular and Cercignani-Lampis BCs) affect the Kramer's problem.

\begin{table}
		\caption{ The VSCs $\bar{\zeta}$ for HS, VHS with $\omega=0.81$, and Maxwell molecules under the diffuse-specular BC with different TMACs. LBE results of the present paper, Siewert~\cite{siewert2003linearized}, and Wakabayashi \textit{et al.}~\cite{wakabayashi1996numerical} are denoted by  a, b, and c, respectively. }
	\label{table_DF_hardsphere}
	\centering

	\begin{tabular}{ccccc|ccccc}
		\hline
		$\alpha_M$
		& 	&  HS & VHS  &   Max.   & $\alpha_M$	& 	&  HS & VHS  &   Max.\\
\hline
		$0.1$
		&a	& 17.04836  & 17.09319  & 17.12847  & $0.6$ &a	&  2.215672  & 2.245067   & 2.267942  \\
		&b	& 17.04780  & --  & --    &&c	&2.214780 & --  & --  \\
		&c	& 17.00580  & --  & --    &&b	& 2.209300  & --  & -- \\

		$0.2$
		&a	& 8.173130   & 8.214580  & 8.247129 & $0.7$ &a	&  1.781936  & 1.808636   & 1.829366  \\
		&b	&8.172480  & --  & --        &&b	&1.780980  & --  & --     \\
		&c	&8.152400  & --  & --        &&c	& 1.776600  & --  & -- \\

		$0.3$
		&a	& 5.206345  & 5.244574   & 5.274524   & $0.8$ &a	&  1.453926  & 1.478044   & 1.496725 \\
		&b	&5.205630  & --  & --  & &b	& 1.452920  & --  & --     \\
		&c	& 5.192800  & --  & -- & &c	& 1.449400  & --  & --    \\

		$0.4$
		&a	& 3.716862 & 3.752014   & 3.779490   &$0.9$&a	&  1.196466  & 1.218108   & 1.234829\\
		&b	&3.716090  & --  & --   &&b	& 1.195400  & --  & --   \\
		&c	& 3.706900  & --  & -- &&c	& 1.192500  & --  & --   \\

		$0.5$
		&a	&  2.818444  & 2.850653   & 2.875774  &		$1.0$ &a	& 0.988451   & 1.007717   & 1.022560  \\
		&b	&2.817610  & --  & --  &		&b	& 0.987328  & --  & --  \\
		&c	& 2.810700  & --  & -- &  	&c	& 0.984900 & --  & --     \\
\hline
	\end{tabular}\par

\end{table}

\subsubsection{The influences of intermolecular potential and gas-kinetic BC}


Table~\ref{table_DF_hardsphere} tabulates the VSCs obtained from the LBE for HS, variable hard-sphere (VHS) with $\omega=0.81$, and Maxwell molecules, when the diffuse-specular BC of different TMACs is used. Results of Wakabayashi \textit{et al.}~\cite{wakabayashi1996numerical} using a discrete velocity method and Siewert~\cite{siewert2003linearized} using a polynomial expansion technique to solve the LBE for HS molecules are also listed for comparison. It is noticed that the three groups of data agree well with each other, especially the relative difference between our results and those of Siewert~\cite{siewert2003linearized} is less than $10^{-4}$. As expected, the VSC increases as the TMAC decreases. Also, the VSC is insensitive to the intermolecular potential, which only slightly increases with the viscosity index $\omega$, where the relative difference between HS and Maxwell molecules is less than $4\%$. This results confirm the statement in previous studies~\cite{sharipov1998data,sharipov2011data,loyalka1975some}.

\begin{table}[t]
		\caption{ The VSC for HS, VHS with $\omega=0.81$, and Maxwell molecules under Cercignani-Lampis BC with different effective TMAC $\alpha_t$ and energy accommodation coefficient $\alpha_n$.}
	\label{table_CL}
	\centering

	\begin{tabular}{cccccc}
\hline
		$\alpha_t$	& $\omega$
		&  $\alpha_n=0.25$ & $ 0.5$  &    $0.75$ &    $ 1$\\
\hline
		$0.25$
		& 0.5	& 6.365427  & 6.343336   & 6.324267 &  6.307321  \\
		&0.81	& 6.400178  & 6.372316   & 6.347638 &  6.325202  \\
		&1	    & 6.426786  & 6.394845   & 6.366237 &  6.339971  \\
		$0.5$
		& 0.5	& 2.799516   & 2.785158   & 2.772602 & 2.761338  \\
		&0.81	& 2.829277   & 2.811279   & 2.795092 & 2.780207  \\
		&1		& 2.851688   & 2.831142   & 2.812430 &  2.795028  \\

		$0.75$
		& 0.5	& 1.598122   & 1.591127   & 1.584932  & 1.579323 \\
		&0.81	& 1.622629   & 1.613906   & 1.605945  & 1.598540 \\
		&1		& 1.641138   & 1.631215   & 1.622031 &  1.613380  \\
		$1.0$
		& 0.5	& 0.988451   & 0.988451   & 0.988451 & 0.988451 \\
		&0.81	& 1.007717   & 1.007717   & 1.007717 & 1.007717 \\
		&1		& 1.022560   & 1.022560   & 1.022560 &  1.022560  \\

		$1.25$
		& 0.5	& 0.615670   & 0.622315   & 0.628343 &  0.633906 \\
		&0.81	& 0.629985   & 0.638188   & 0.645891  & 0.653221 \\
		&1		& 0.641382   & 0.650657   & 0.659514 &  0.668067  \\

		$1.5$
		& 0.5	& 0.361248   & 0.374217   & 0.386121 & 0.397213 \\
		&0.81	& 0.371198   & 0.387115   & 0.402275 &0.416866 \\
		&1		& 0.379386   & 0.397328   & 0.414721 & 0.431729  \\

		$1.75$
		& 0.5	& 0.174178   & 0.193187   & 0.210840   & 0.227456\\
		&0.81	& 0.180631   & 0.203809   & 0.226193  & 0.247988 \\
		&1.0	& 0.185886   & 0.211919   & 0.237535 &  0.262897  \\

		$2$
		& 0.5	& 0.028851   & 0.053665   &0.076984 & 0.099153\\
		&0.81	& 0.032881   & 0.062901   &0.092298  & 0.121255 \\
		&1.0	& 0.035527   & 0.069105   &0.102637 &  0.136246  \\
		\hline
	\end{tabular}\par

\end{table}

In order to study the Kramer's problem with a more sophisticated gas-surface interaction, the LBE is then solved with the Cercignani-Lampis BC. Results are summarized in Table~\ref{table_CL}, for the effective TMAC  $\alpha_t\in[0.25, 2]$ and the energy accommodation coefficient $\alpha_n\in[0.25,1]$. When the value of $\alpha_n$ and the intermolecular potential are fixed, the VSC increases rapidly when $\alpha_t$ decreases, which is consistent with that in the diffuse-specular BC. The additional free parameter $\alpha_n$ in Cercignani-Lampis BC introduces new interesting results. When $\alpha_t<1$, for a fixed $\alpha_t$ and intermolecular potential, the VSC decreases slightly as $\alpha_n$ increases, where the maximum drop in the VSC is less than $2\%$. When $\alpha_t=1$, the Cercignani-Lampis BC is reduced to the fully diffuse one in this problem, and the VSC does not vary with $\alpha_n$. When $\alpha_t>1$, the variation of VSC on $\alpha_n$ reverses when compared to that of $\alpha_t<1$; and it is strongly influenced by $\alpha_n$, especially when $\alpha_t$ is large. For instance, for HS molecules at $\alpha_t=2$, the VSC is increased by more than three times when $\alpha_n$ changes from $0.25$ to $1$. For fixed $\alpha_n$ and $\alpha_t$, the change in the VSC is insensitive to the intermolecular potentials when $\alpha_t\lesssim1.75$. However, when $\alpha_t$ is close to two (i.e. the ``backward'' scattering), the influence of the intermolecular potential becomes considerable. For example, when $\alpha_t=2$ and $\alpha_n=1$, Maxwell molecules have a VSC that is about 37\% higher than that for HS molecules.

\subsubsection{The viscous slip coefficient as a function of the effective TMAC}

The variation of the VSC with respect to the effective TMAC $\alpha$ (for diffuse-specular and Cercignani-Lampis BCs, $\alpha=\alpha_M$ and $\alpha_t$, respectively) could be generalized to some simple expressions. By considering the VSCs at the two limit ends of $\alpha=1$ and $\alpha\rightarrow 0$, the VSC is fitted by Eq.~\eqref{eq:loyalka} for the diffuse-specular BC. For Cercignani-Lampis BC, Sharipov proposed a similar equation which is a linear combination of the VSCs at $\alpha_t=1$ and $\alpha_t=2$~\cite{Sharipov2003CL}. However, the estimation shows a large error when $\alpha_t\rightarrow 0$. Here we construct a more accurate expression for the VSC with respect to the effective TMAC $\alpha$.

\begin{figure}
 	\begin{centering}
 		\includegraphics[width=1.1\textwidth]{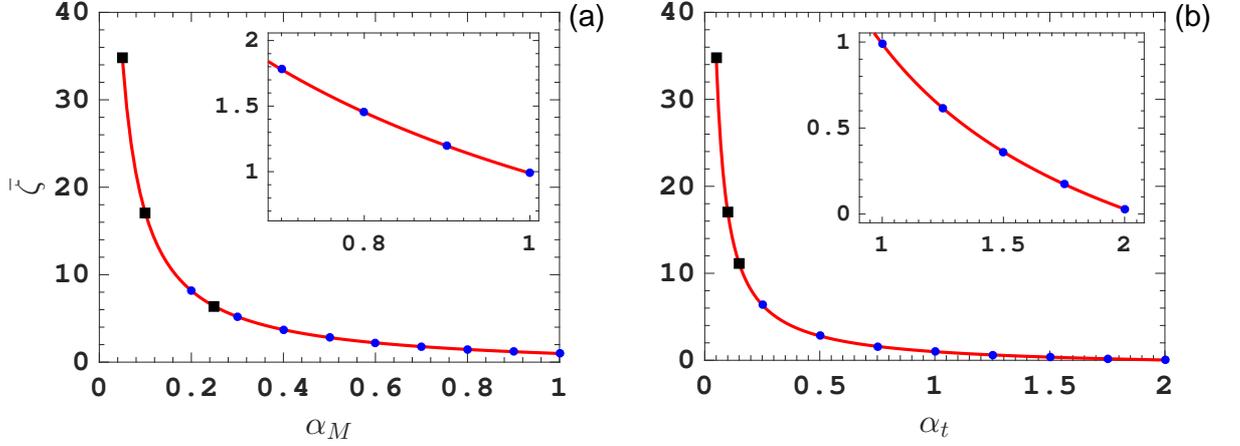}
 		\par\end{centering}
 	\caption{ The VSC as a function of the effective TMAC $\alpha$ for HS molecules, when the (a) diffuse-specular BC and (b) Cercignani-Lampis BC with $\alpha_n=0.25$  are used. Solid lines: the numerical fitting of Eq.~\eqref{CL_fit} using the data from the LBE solutions (circles). Squares: the data from the LBE but no used for fitting. Note that other values of $\alpha_n$ and other types of intermolecular potentials show a similar behavior.
 	}
 	\label{fig:slipCoe_fit}
 \end{figure}

We find from Tables~\ref{table_DF_hardsphere} and~\ref{table_CL} that the VSC can be fitted by a general function as the one used by Lilley \& Sader for both diffuse-specular and Cercignani-Lampis BCs~\cite{lilley2008velocity}:
 \begin{equation}\label{CL_fit}
\bar{\zeta}(\alpha)=\frac{a}{\alpha}-b\alpha-c,
\end{equation}
where the fitting coefficients $a,~b$, and $c$ are shown in
Table~\ref{table:slipcoe_fit} for typical inverse power-law intermolecular potentials. Fig.~\ref{fig:slipCoe_fit} shows that the fitted curve (constructed from the data when $\alpha\geq0.2$) can accurately predict the VSC even in the limit $\alpha\rightarrow 0$. For instance,  when the TMAC is 0.05, relative differences between the fitted VSC and the LBE solutions are less than $0.1\%$ for both diffuse-specular and Cercignani-Lampis BCs.

 \begin{table}[t]
 		\caption{Fitting coefficients in Eq.~\eqref{CL_fit} for HS, VHS with $\omega=0.81$, and Maxwell molecules under the diffuse-specular and Cercignani-Lampis BCs. }
 	\label{table:slipcoe_fit}
 	\centering

 	\begin{tabular}{ccccccc}
\hline
 BC &		$\alpha_n$ & $\omega$
 		&  $a$ & $b\cdot 10$  &   $c$   \\
\hline

 Diffuse-specular &	 n/a	& 0.5
 		& 1.773  & 1.1660  &  0.6687   \\
 	& n/a	& 0.81
 		& 1.773  & 1.4270  &  0.6238   \\
 	& n/a	& 1
 		& 1.773  & 1.6370  &  0.5885   \\
 Cercignani-Lampis &		$0.25$ & 0.5
 		& 1.774  & 0.7266  &  0.7127   \\
 	&	& 0.81
 		& 1.775  & 0.8889  &  0.6781   \\
 	&	& 1
 		& 1.776  & 1.0130  &  0.6516   \\

  Cercignani-Lampis &		$0.5$ & 0.5
 		& 1.773  & 0.4772 &  0.7367 \\
 	&	& 0.81
 		& 1.774  & 0.5867  &  0.7069   \\
 	&	& 1
 		& 1.774  & 0.5742  &  0.6837  \\

  Cercignani-Lampis &		$0.75$ & 0.5
 		& 1.772   & 0.2434  &  0.7597   \\
 	&	& 0.81
 		& 1.773  & 0.2915  &  0.7358  \\
 	&	& 1
 		& 1.773  & 0.3373  &  0.7164  \\
  Cercignani-Lampis &		$1.0$ &0.5
 		& 1.772   &0.0218    & 0.7818 \\
 	&	& 0.81
 		& 1.766  & 0.0543  &  0.7370   \\
 	&	& 1
 		& 1.772  & 0.0000  &  0.7499  \\
 		\hline
 	\end{tabular}\par
 
 \end{table}

\subsection{The Knudsen layer function }\label{sec:knudsenlayer}

%

\subsubsection{The influence of the intermolecular potential}

\begin{figure}
	\centering
	\includegraphics[width=0.8\textwidth]{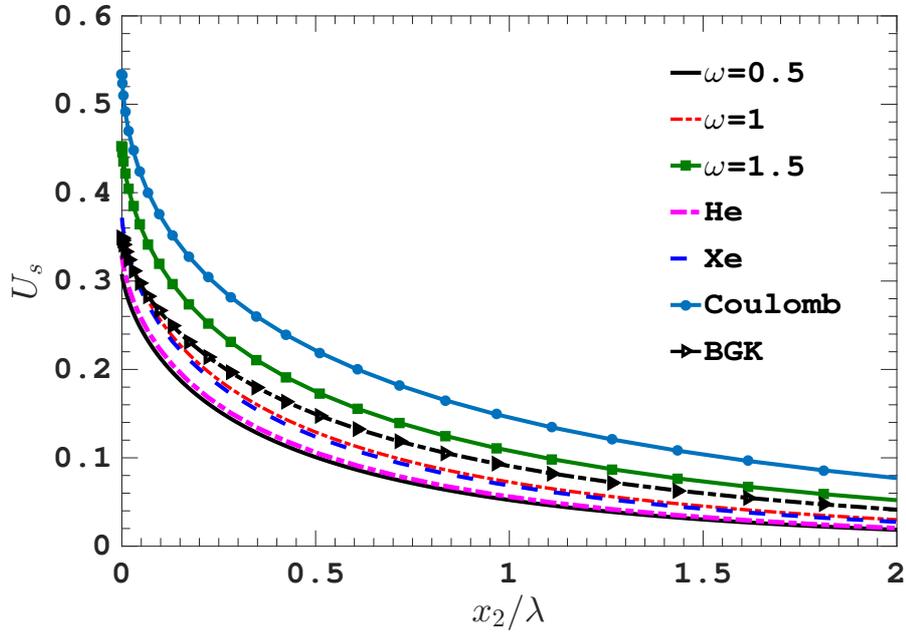}
	\caption{ KLFs from the LBE solutions for the inverse power-law potentials, the Lennard-Jones potentials of Helium and Xenon when the gas temperature is $300$K, and the shielded Coulomb potential of charged molecules when $\epsilon=k_BT_w$. The result from BGK equation is also included for comparison. The diffuse BC is used.  }
	\label{fig:vel_def_omega}
\end{figure}

Figure~\ref{fig:vel_def_omega} illustrates the KLFs obtained from the LBE with the  diffuse BC, for the inverse power-law potential with various values of viscosity index $\omega$, the Lennard-Jones potential of helium and xenon, and the shielded Coulomb potential. It is found that, for the inverse power-law potential, the KLF increases with the viscosity index in the whole Knudsen layer. For the Lennard-Jones potential, the KLF of xenon molecules is larger than that of helium, but the results of both helium and xenon lie between those of HS and Maxwell molecules. This is comprehensible because the effective viscosity indexes of helium and xenon at a temperature of $300$K are 0.66 and 0.85~\cite{Bird1994}, respectively.  The KLF predicted by the BGK is even larger than that from the Maxwell molecules, but is smaller than that of $\omega=1.5$ where the gas molecules interact with soft potentials. The shielded Coulomb potential has the largest KLF, since its effective viscosity is close to 2.5~\cite{CE}.

Thus, contrary to the VSC whose value is insensitive to the intermolecular potential, the KLF is strongly affected by the intermolecular potential. That is, when the effective viscosity index increases, (i) the value of the KLF increases, and (ii) the KLF decays more slowly, or equivalently, the Knudsen layer becomes wider. For example, at the solid surface, the relative difference between KLFs of Maxwell and HS molecules is approximate $20\%$, and that between the shielded Coulomb and HS potentials reaches 60\%. Relative differences at distances one to two mean free path away from the solid surface are even larger, say, when $x_2/\lambda=2$ the value of the KLF of the shielded Coulomb potential is about 4 times of that of HS potential. On the other hand, when $U_s$ is decreased to 0.01 of its value at the solid surface, the corresponding distances to the solid surface for the HS and Maxwell molecules are about $2.7\lambda$ and $3.5\lambda$, respectively.

\subsubsection{The influence of the gas-kinetic BC}

\begin{figure}[t]
	\centering
	\includegraphics[width=1.1\textwidth]{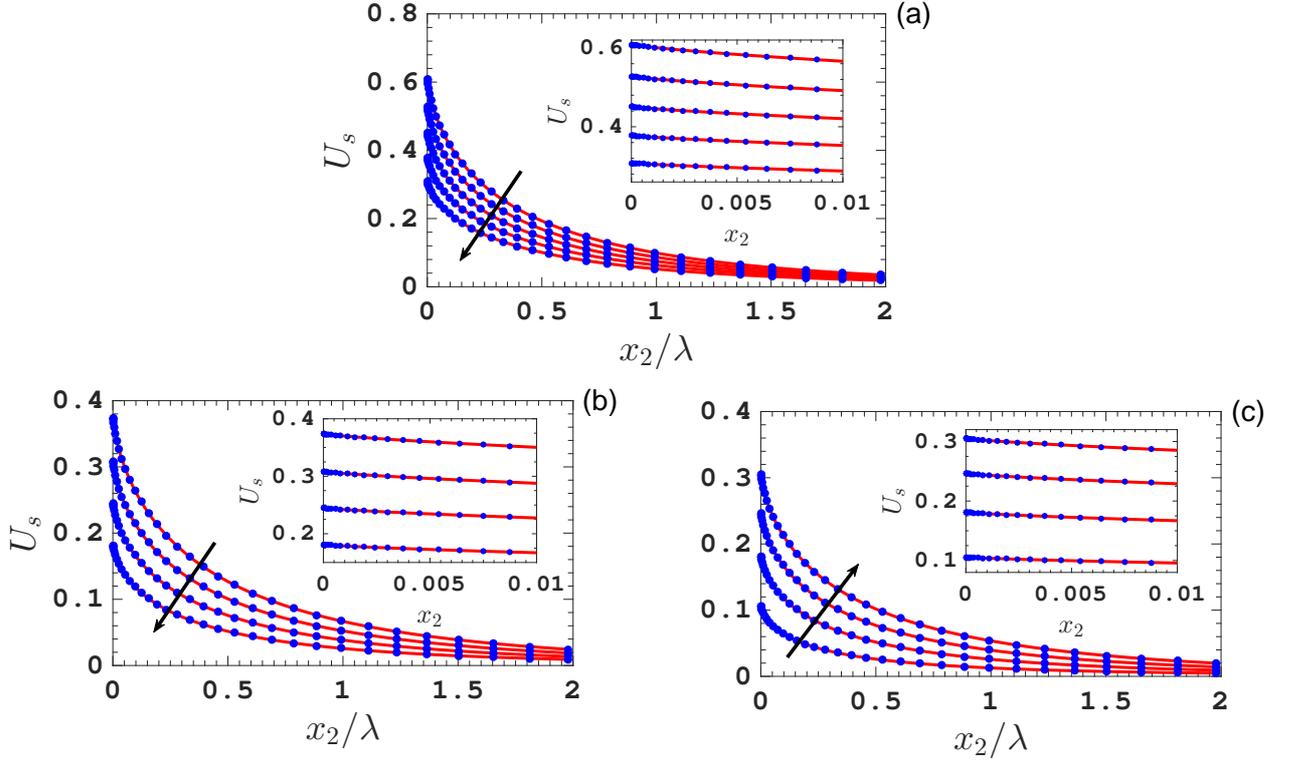}
	\caption{The KLF for HS molecules. (a) The diffuse-specular BC. Along the arrow,   $\alpha_M=0.2,0.4,0.6,0.8,1$. (b) The Cercignani-Lampis BC with $\alpha_n=0.25$. Along the arrow, $\alpha_t=0.5,~1,~1.5$ and 2. (c) The Cercignani-Lampis BC with $\alpha_t=2$. Along the arrow, $\alpha_n=0.25,~0.5,~0.75$ and 1. Dots: LBE solutions. Solid lines: fitted curves using Eq.~\eqref{eq:vel_fit} with $M,N=2$. Insets: zoomed regions in the vicinity of the solid surface. }
	\label{fig:vel_def_cl_max}
\end{figure}

For the sake of clarity, we only focus on  HS molecules. The diffuse-specular BC is first considered in Fig.~\ref{fig:vel_def_cl_max}(a). It is found that the KLF increases as $\alpha_M$ decreases. For a fixed $\alpha_M$, the KLF decreases rapidly when moving away from the solid surface, say, its value decays by roughly $85\%$ of the value on the solid surface when $x_2$ is about one mean free path away from the solid surface.

Typical KLFs under Cercignani-Lampis BC are included in Figs.~\ref{fig:vel_def_cl_max}(b) and~(c). When $\alpha_n$ is fixed, for example, for $\alpha_n=0.25$, the KLF decreases as $\alpha_t$ increases. The relative reduction in $U_s(0)$ is about $40\%$ when $\alpha_t$ rises from 0.25 to 1. However, the variation of the KLF with respect to $\alpha_t$ becomes weaken as $\alpha_n$ increases, such that at $\alpha_n=1$ the reduction in $U_s(0)$ with $\alpha_t$ falls below $3\%$ (this is not visualized here but can be deducted from Table~\ref{table_cl_coe} below). When $\alpha_t$ ($\neq1$) is fixed, the influence of $\alpha_n$ on the KLF becomes larger when $\alpha_n$ increases. And the greater the TMAC $\alpha_t$ exceeds 1, the more pronounced the change in the KLF with $\alpha_n$. As an example, when $\alpha_t=2$, the KLF is increased by three times, as the $\alpha_n$ varies from 0.25 to 1, see Fig.~\ref{fig:vel_def_cl_max}(c).


\begin{table}[pt]
		\caption{ Fitted coefficients  corresponding to Eq.~\eqref{eq:vel_fit} with $M,N=2$ for the KLF obtained from the LBE with HS, VHS with $\omega=0.81$, and Maxwell molecules, when the diffuse-specular BC is used.}
	\label{table_df_coe}
	\centering

	\begin{tabular}{ccccccccccl}
\hline
		$\omega$	& $\alpha_M$
		&  $c_{0,0}$ & $c_{0,1}$  &   $c_{0,2}$ &   $c_{1,0}$ &   $c_{1,1}$ &   $c_{1,2}\cdot10$ &   $c_{2,0}$ &   $c_{2,1}$ &   $c_{2,2}\cdot 10^{2}$   \\
\hline
		$0.5$
		& 0.1 & 0.6502  & 1.2720  & -0.3624& 1.4640  & 0.9950  & -0.6074  & -2.0090  &  0.1593 &0.2246   \\
		& 0.2 & 0.6084  & 1.1760  & -0.3257& 1.3260  & 0.9093  & -0.5358  & -1.8360  & 0.1410 &0.1967   \\
		& 0.3 & 0.5675  &1.0830   & -0.2915& 1.1970  & 0.8283  & -0.4699  & -1.6720  & 0.1241 &0.1712   \\
		& 0.4 & 0.5277  &0.9941   & -0.2598& 1.0760  & 0.7517  & -0.4095  &-1.5170   & 0.1085 & 0.1479  \\
		& 0.5  & 0.4889 &0.9091   &-0.2302 & 0.9630  & 0.6792  &  -0.3543 & -1.3710 & 0.0943 &0.1267  \\
		& 0.6  & 0.4509 &0.8277   &-0.2029 & 0.8570  & 0.6108  &  -0.3040 & -1.2330  & 0.0813 &0.1075  \\
		& 0.7  & 0.4139 & 0.7498  &-0.1776 & 0.7582  &0.5463   &  -0.2585 & -1.1030  & 0.0695 &0.0902  \\
		& 0.8  & 0.3778 & 0.6752  &-0.1543 & 0.6661  &0.4856   &  -0.2173 & -0.9805 & 0.0588 &0.0748  \\
	    & 0.9  & 0.3424 & 0.6038  &-0.1328 & 0.5805  &0.4284   &  -0.1805 & -0.8652 & 0.0492 &0.0610 \\
		& 1.0  & 0.3079 & 0.5356  &-0.1132 & 0.5012  & 0.3746 & -0.1480   & -0.7568 &0.0406 &0.0489  \\
		$0.81$
		& 0.1 & 0.7436 & 1.6500   & -0.6657& 2.3940  &  1.4840 & -1.3430 &-3.0070 & 0.3430 &0.5216 \\
		& 0.2 & 0.6935 & 1.5130   & -0.5949& 2.1550  &  1.3470 & -1.1840 &-2.7270 & 0.3030 &0.4571 \\
		& 0.3 & 0.6450  & 1.3840   & -0.5294& 1.9330  &  1.2180 & -1.0380 &-2.4640& 0.2664 & 0.3983 \\
	    & 0.4 & 0.5980  & 1.2610   & -0.4691& 1.7270  &  1.0980 & -0.9046 &-2.2190& 0.2329 & 0.3449 \\
		& 0.5 & 0.5523 & 1.1450   & -0.4135& 1.5360  & 0.9858 &-0.7836  &-1.9890 & 0.2024 &0.2966 \\
	    & 0.6 & 0.5081 & 1.0360   & -0.3624& 1.3590  & 0.8806  &-0.6739  &-1.7750 & 0.1747 &0.2531 \\
	    & 0.7 & 0.4651 & 0.9315   & -0.3155& 1.1940  & 0.7823  &-0.5747  &-1.5750 & 0.1495 &0.2139 \\
	    & 0.8 & 0.4233 & 0.8331   & -0.2727& 1.0430  &0.6907  &-0.4855  &-1.3890 & 0.1269 &0.1789 \\
	    & 0.9 & 0.3827 &0.7400    & -0.2336& 0.9035  &0.6052  &-0.4056  &-1.2160& 0.1065 &0.1478\\
		& 1.0 & 0.3433 & 0.6519   & -0.1981& 0.7753  &0.5257  &-0.3345   &-1.0550  &0.0883 &0.1203 \\
		$1$
	    & 0.1 &0.8123  &  1.9930   &-1.0180 & 3.3540  & 1.9460 &-2.3230   &-4.0160 &  0.5808 &0.9405  \\
		& 0.2 &0.7559  & 1.8180    &-0.9055 & 3.0050  & 1.7580 &-2.0430   &-3.6190 &  0.5118 &0.8235  \\
	    & 0.3 &0.7016  & 1.6530    &-0.8023 & 2.6820  &1.5820 &-1.7880   &-3.2510 &  0.4490 &0.7174 \\
		& 0.4 &0.6491  & 1.4980    &-0.7076 & 2.3840  &1.4190 &-1.5570   &-2.9100 &  0.3918 &0.6214 \\
		& 0.5 &0.5983  & 1.3530   &-0.6210 & 2.1090  & 1.2670 &-1.3470 &-2.5930 & 0.3400 &0.5349   \\
		& 0.6 &0.5493  &1.2170     &-0.5419 & 1.8570  & 1.1260 &-1.1580 &-2.3000  & 0.2931 &0.4571 \\
		& 0.7 &0.5019  &1.0880     &-0.4699 & 1.6240  & 0.9955 &-0.9880 &-2.0290  & 0.2508 &0.3874\\
		& 0.8 &0.4560   &0.9682    &-0.4044 & 1.4120  & 0.8745 &-0.8354 &-1.7790  & 0.2127 &0.3252\\
		& 0.9 &0.4116  &0.8553    &-0.3450  & 1.2170  & 0.7625 &-0.6991 &-1.5480  & 0.1787 &0.2700\\
		& 1.0 &0.3685  &  0.7495  &-0.2914 & 1.0390  & 0.6590 &-0.9010 &-1.3350 & 0.1483  &0.2212   \\
		\hline
	\end{tabular}\par
\end{table}



\begin{table}[thbp]
	\centering
	\caption{Fitting coefficients corresponding to Eq.~\eqref{eq:vel_fit} with $M,N=2$ for the LBE solutions of the KLF, when  HS molecules and  Cercignani-Lampis BC are used. Since the KLF is independent of $\alpha_n$ when $\alpha_t=1$, only the fitting coefficients for $\alpha_t=1$ and $\alpha_n=0.25$ are tabulated. }
\label{table_cl_coe}
	\begin{tabular}{ccccccccccl}
\hline
		$\alpha_n$	& $\alpha_t$
		&  $c_{0,0}$ & $c_{0,1}$  &   $c_{0,2}$ &   $c_{1,0}$ &   $c_{1,1}$ &   $c_{1,2}\times 10$ &   $c_{2,0}$ &   $c_{2,1}\times 10$ &  $c_{2,2}\times 10^{3}$   \\
\hline
		$0.25$
		& 0.25 & 0.4756    & 0.7342  & -0.1390  & 0.5628 & 0.4537  & -0.2119  & -0.9529  & 0.5551 & $0.7804$   \\
		& 0.5  & 0.4174   & 0.6648  & -0.1299  & 0.5416 & 0.4265  & -0.1888  & -0.8847  & 0.5012 &$0.6781$   \\
		& 0.75 & 0.3616   & 0.5988  & -0.1214  &0.5210  & 0.4001  & -0.1676  & -0.8195  & 0.4520 & $0.5814$   \\
		& 1    & 0.3079  &0.5356   & -0.1132  &0.5012  & 0.3746  &  -0.1476 & -0.7568  & 0.4056 &$0.4893$  \\
		& 1.25 & 0.2559   &0.4746   & -0.1053  & 0.4823 &0.3501   & -0.1283  &  -0.6965 &0.3607  &$0.4004$ \\
		& 1.5  & 0.2051  &0.4151   & -0.0975  & 0.4643 &0.3265   & -0.1089  &  -0.6379 &0.3156  &$0.3134$ \\
		& 1.75 & 0.1551  &0.3563   & -0.0897  & 0.4473 &0.3037   & -0.0891  &  -0.5806 &0.2687  &$0.2270$ \\
		& 2    & 0.1056  & 0.2980  &-0.0816   & 0.4316 &0.2818   & -0.0683  &  -0.5245 &0.2188  &$0.1405$ \\
		$0.5$
		& 0.25 & 0.4080 &  0.6424  & -0.1285  &  0.5242   & 0.4152 &  -0.1848 & -0.8585  & 0.4961 &$0.6048$\\
		& 0.5  & 0.3735 & 0.6048   & -0.1231  &  0.5159   & 0.4011 &  -0.1714 & -0.8228  & 0.4635 &$0.5872$ \\
		& 0.75 &0.3403  & 0.5694   & -0.1180  &  0.5084   & 0.3877 &  -0.1592 & -0.7891  & 0.4338 &$0.5372$ \\
		& 1.25 &  0.2762& 0.5023 & -0.1085  & 0.4942    & 0.3619 &-0.1363   & -0.7252  & 0.3781 &$0.4524$ \\
		& 1.5  & 0.2446   &0.4689  &  -0.1037 & 0.4871    &0.3491  &-0.1247   &-0.6934   & 0.3498 &$0.3951$ \\
		& 1.75 & 0.2130  & 0.4345 &  -0.0987 &  0.4797   &0.3363  &-0.1124   &-0.6608   & 0.3198 &$0.3462$ \\
		& 2    & 0.1810   & 0.3987 & -0.0935  &   0.4719  &0.3232  &-0.0992   &-0.6271   &0.2873  &$0.2953$ \\
		$0.75$
		& 0.25 &0.3545   & 0.5836  &-0.1203  &  0.5099 &0.3926 &-0.1647 &-0.8019 &  0.4479 &$0.5554$   \\
		& 0.5  &0.3383   & 0.5658  &-0.1177  &  0.5061 & 0.3861 &-0.1582 &-0.7852 &  0.4319 &$0.5304$   \\
		& 0.75 &0.3229   & 0.5502  &-0.1154  &  0.5034 & 0.3802 &-0.1526 &-0.7705 &  0.4182 &$0.5090$   \\
		& 1.25&0.2931   & 0.5211  &-0.1111   & 0.499  & 0.3691 &-0.1426 &-0.7433 &  0.3932 &$0.4698$   \\
		& 1.5 &0.2780   & 0.5058  &-0.1088  & 0.4964  & 0.3634 &-0.1372 &-0.7290 &  0.3798 &$0.4489$   \\
		& 1.75&0.2625   & 0.4888  &-0.1063  & 0.4929  & 0.3572 &-0.1310 &-0.7130 &  0.3645 &$0.4251$   \\

		& 2   & 0.2462  & 0.4696  &-0.1035 & 0.4884   & 0.3504 &-0.1238 &-0.6951 &  0.3466  &$0.3979$   \\

		$1$
		& 0.25&0.3093   & 0.5405   &-0.1138 & 0.504 & 0.3762 &-0.1499 &-0.7617 &  0.4109 &0.4980   \\
		& 0.5 &0.3083   & 0.5370   &-0.1134  & 0.5020 & 0.3751 &-0.1483 &-0.7582 &  0.4071  &0.4918   \\
		& 0.75&0.3080   & 0.5357   &-0.1132  & 0.5013 & 0.3747 &-0.1477 &-0.7570 &  0.4058  &0.4896  \\
		& 1.25&0.3079   & 0.5354   &-0.1132  & 0.5011 & 0.3746 &-0.1475 &-0.7567 &  0.4055 &0.4890   \\
		& 1.5 &0.3075   & 0.5342  &-0.1130  & 0.5004  & 0.3742 &-0.1470  &-0.7555  &  0.4042  &0.4869   \\
		& 1.75 & 0.3066  & 0.5311  &-0.1127 & 0.4987  & 0.3732 &-0.1455 &-0.7524  &  0.4008  &0.4813   \\
		& 2   & 0.3050  & 0.5255  &-0.1120 & 0.4955   & 0.3715 &-0.1430  & -0.7469 & 0.3949   &0.4714   \\
		\hline
	\end{tabular}\par

\end{table}


\subsubsection{Fitting the Knudsen layer function and the singularity of velocity gradient}

The KLF is essential not only in determining the nonlinear constitution in the Knudsen layer~\cite{lockerby2005capturing}, but also in defining the singularity of the velocity gradient near the solid surface. In a recent work based on the BGK model, Jiang \& Luo~\cite{Jiang2016JCP} have rigorously shown that the velocity near the solid surface can be described by Eq.~\eqref{eq:vel_fit}, whose gradient possesses a logarithmic divergence. However, in their work, it was only numerically demonstrated that the first four leading terms of Eq.~\eqref{eq:vel_fit} can capture the velocity profile in an extremely small interval $0\leq{}x_2\leq1.5\times 10^{-7}$.

In this work, based on the highly accurate results of the LBE, surprisingly, we find that the entire KLF can be described by Eq.~\eqref{eq:vel_fit}, provided that more high-order terms are included. The associated fitting coefficients in Eq.~\eqref{eq:vel_fit} with $M, N=2$  for the diffuse-specular and Cercignani-Lampis BCs are tabulated in Tables~\ref{table_df_coe} and~\ref{table_cl_coe}, respectively. We note that when $\alpha_n$ is fixed, the absolute value of the fitting coefficient decreases as $\alpha_t$ increases. Meanwhile, the dependency of each fitting coefficient on $\alpha_t$ becomes weaker and weaker as $\alpha_n$ increases. For instance, when $\alpha_n$ is increased from 0.25 to 1, the maximum relative difference in $c_{0,0}$ for different $\alpha_t$ is reduced from $350\%$ to $2\%$. From the insets in Fig.~\ref{fig:vel_def_cl_max}, we observe that the fitted curves agree quite well with the numerical results. Note that Eq.~\eqref{eq:vel_fit} with $M, N=2$ can also describe the KLF very well when the distance to the solid surface reaches $10\lambda$.

\begin{figure}
	\centering
	\includegraphics[width=0.8\textwidth]{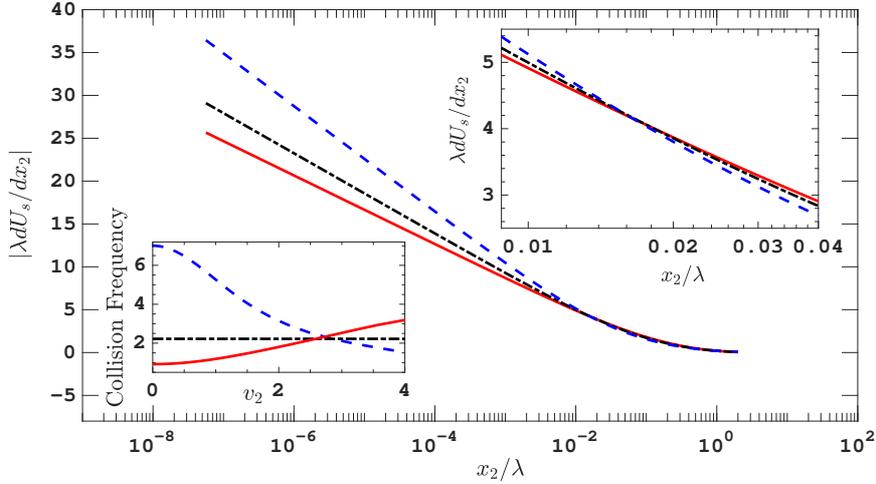}
	\caption{The absolute value of the velocity gradient $|\lambda dU_s/dx_2|$ for HS (solid line), Maxwell (dash-dot line), and soft-potential with $\omega=1.5$ (dash line) molecules, when the diffuse boundary is used. Insets are the zoomed velocity gradient and the equilibrium collision frequency $\nu_{eq}(0,v_2,0)$ normalized by the rarefaction parameter $\delta$.   }
	\label{fig:vel_grad}
\end{figure}

Next, the singularity of the velocity gradient in the vicinity of the solid surface is investigated through the deviation of Eq.~\eqref{eq:vel_fit} with respect to $x_2$. This singularity is dominated by the term with $n=0$ and $m=1$ in Eq.~\eqref{eq:vel_fit}, that is, the velocity gradient near the solid surface is $c_{0,1}\ln{}x_2$~\cite{takata2013singular,Jiang2016JCP}. 
From Table~\ref{table_df_coe}, it is found that for a fixed TMAC, $c_{0,1}$ increases with the viscosity index, indicating that the rate of divergence is faster for the gas molecules with a larger value of the viscosity index, see Fig.~\ref{fig:vel_grad}. However, this trend reverse at $x_2\approx0.015\lambda$. This behavior is somehow related to the variation of the equilibrium collision frequency $\nu_{eq}$. From the left inset in Fig.~\ref{fig:vel_grad} we see that, when the rarefaction parameter $\delta$ is fixed, $\nu_{eq}(0,0,0)$ increases with the viscosity index $\omega$, which means that the collision frequency is larger for larger values of $\omega$, so that the gas approaches to the equilibrium quicker and hence the velocity defect decreases faster. Similarly, the velocity gradient at $x_2>0.015\lambda$ seems to be proportional to $\nu_{eq}(0,v_2>3,0)$. It should be noted that, however, this explanation is phenomenological; one may resort to the rigorous mathematical analysis to have a deep understanding~\cite{takata2013singular,takata2011JFM}. When the intermolecular potential is fixed, a smaller effective TMAC will produce a larger velocity gradient near the solid surface, see Tables~\ref{table_df_coe} and~\ref{table_cl_coe}.

\begin{figure}[t]
	\centering
	\includegraphics[width=0.8\textwidth]{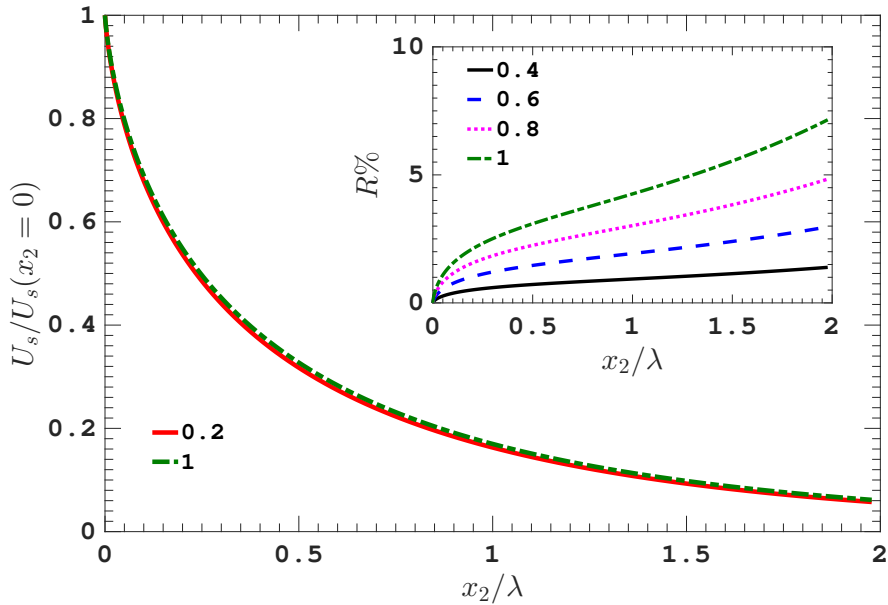}
	\caption{The rescaled KLF $U_s/U_s(x_2=0)$ for  $\alpha_M=0.2$ and 1, when the HS molecules and the diffuse-specular BC are used. For clarity, results at other values of $\alpha_M$ are not shown. Inset: the relative difference ($R\%$) of $U_s/U_s(x_2=0)$ for various $\alpha_M$ compared to that of $\alpha_M=0.2$. }
	\label{fig:vel_def_rescale}
\end{figure}

\begin{figure}[t]
	\centering
	\includegraphics[width=1\textwidth]{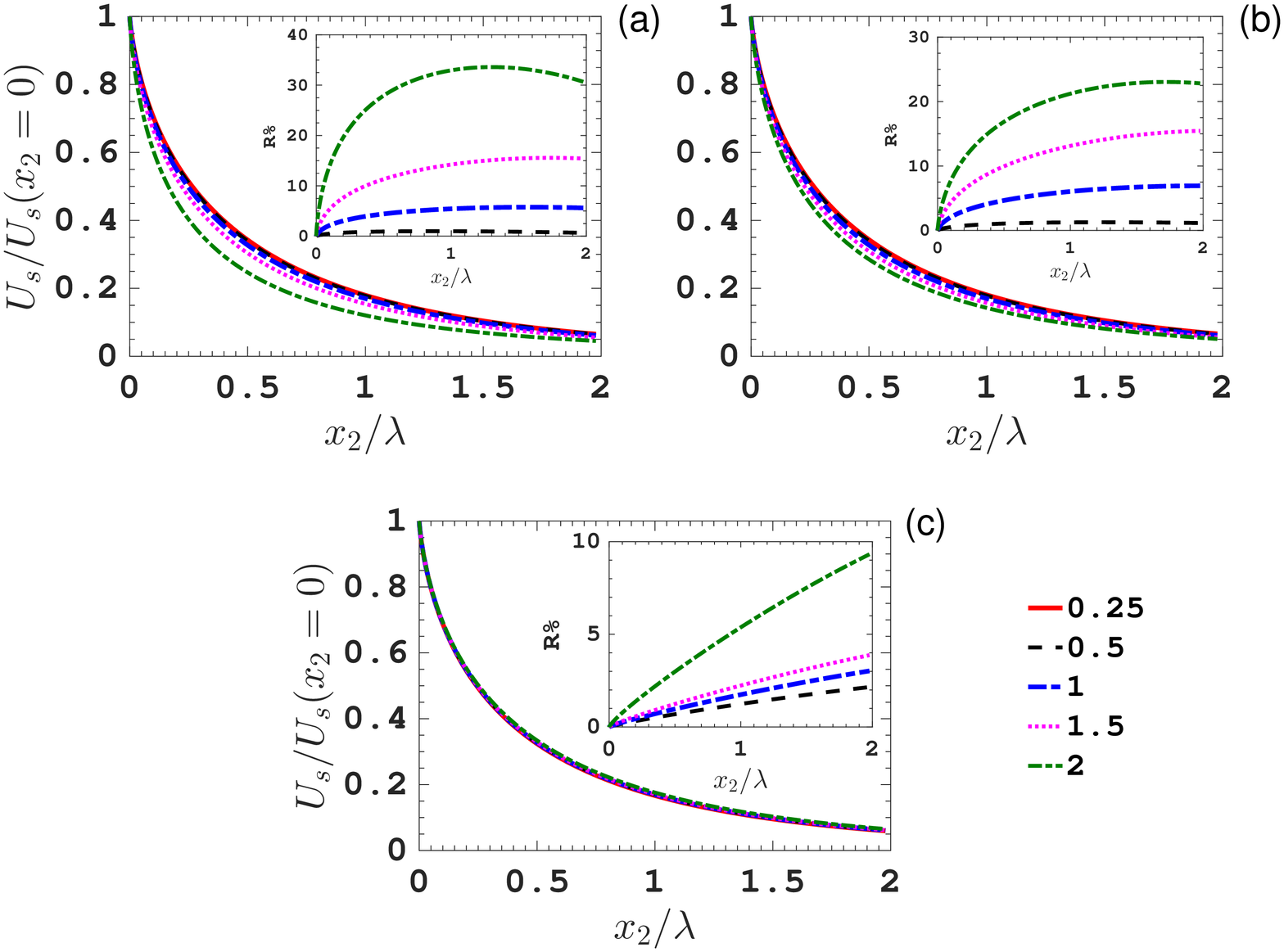}
	\caption{The rescaled KLF $U_s/U_s(x_2=0)$ of HS molecules when (a) $\alpha_n=0.25$, (b) $\alpha_n=0.5$, and (c) $\alpha_n=1$, when the Cercignani-Lampis BC with $\alpha_t=0.25, 0.5, 1.0, 1.5$, and 2.0 are used. Inset: the relative difference ($R\%$) of $U_s/U_s(x_2=0)$ at various TMAC, when compared to that of $\alpha_t=0.25$.   }
	\label{fig:vel_def_rescale_CL}
\end{figure}

\begin{table}[thbp]
	\centering
	\caption{ Fitting coefficients of the rescaled KLF $U_s/U_s(0)$ for inverse power-law potentials with different values of the viscosity index $\omega$, when the diffuse BC is used. }
\label{table_rescale_coe}
	\begin{tabular}{ccccccccccl}
\hline
		$\omega$
		&  $c_{0,0}$ & $c_{0,1}$  &   $c_{0,2}$ &   $c_{1,0}$ &   $c_{1,1}$ &   $c_{1,2}\cdot10$ &   $c_{2,0}$ &   $c_{2,1}$ &   $c_{2,2}\cdot 10^{2}$   \\
\hline
		$0.5$
		& 1.0000 &  1.739 &-0.3677 & 1.628  & 1.217  & -0.4794  & -2.458 &0.1317 &0.1589 \\
		$0.75$
		&1.0000  & 1.864   & -0.5256 &  2.113 &1.462 &-0.8429   &-2.932  &0.2245 &0.2976\\
		$1$
		&1.0000  &  2.034  &-0.7907 & 2.820  & 1.788 &-1.5680 & -3.623 & 0.4024  &0.6003  \\
		$1.25$
       &1.0000  &  2.275  & -1.2390 & 3.864 & 2.217  &-2.9800 & -4.648 & 0.7380   & 1.2300 \\
       	$1.5$
       &1.0000  &   2.807  & -2.5280 & 6.238& 2.821   &-8.4950 & -6.999  & 1.9310   &  4.4710 \\
       \hline
	\end{tabular}\par
\end{table}

\subsubsection{The similarity of the Knudsen layer function}

In the above section, the details of the KLFs under several specific TMACs have been presented, which can serve as benchmark solution. In this section we investigate the similarity in the structure of the Knudsen layer.

We first study the KLF normalized by its value on the solid surface $x_2=0$, when the HS molecules and the diffuse-specular BC are used. Results of other types of molecules are similar. Fig.~\ref{fig:vel_def_rescale} shows the rescaled KLF $U_s/U_s(x_2=0)$ and their relative difference at different TMAC, when compared with that at $\alpha_M=0.2$. We notice that the rescaled KLF for $\alpha_M=0.2$ and 1 almost overlap; as shown in the inset of figure~\ref{fig:vel_def_rescale}, the maximum relative difference among all TMACs is less than $7\%$. Thus, the KLF for diffuse-specular BC possesses a good similarity between different values of TMAC.

For Cercignani-Lampis BC, as can be seen from Fig.~\ref{fig:vel_def_rescale_CL}, when $\alpha_n=1$, the maximum relative discrepancy for all $\alpha_t$ is less than $10\%$. When $\alpha_n$ decreases, however, the deviation of the rescaled KLF between different $\alpha_t$ increases. For instance, when $\alpha_n=0.25$, the maximum relative difference for $\alpha_t=2$ is about $30\%$, as compared with $\alpha_t=0.25$, see the inset in Fig.~\ref{fig:vel_def_rescale_CL}. Nevertheless, it should be noted that, for all $\alpha_n$ with $\alpha_t\le 1 $, the relative difference of the rescaled KLF is less than $7\%$.

Approximately, the KLFs is defined to have similarity if the relative difference of the rescaled KLF for different TMAC is less than $10\%$. Therefore, as shown in Figs.~\ref{fig:vel_def_rescale} and~\ref{fig:vel_def_rescale_CL}, the KLF has the similarity when the diffuse-specular BC and the Cercignani-Lampis BC with $\alpha_n=1$ are considered, in the full range of the effective TMAC; for Cercignani-Lampis BC with other values of $\alpha_n$, the similarity is preserved when $\alpha_t\leq1$. 

Under the diffuse-specular BC, the rescaled KLF can be fitted using Eq.~\eqref{eq:vel_fit}, with the fitting coefficients for different intermolecular potentials tabulated in Table~\ref{table_rescale_coe}. Furthermore, the corresponding KLF on the solid surface $x_2=0$ can be fitted using an exponential function of the effective TMAC $\alpha$ as
\begin{equation}\label{eq:endpoint_fit}
U_s(x_2=0)=c_1\exp(-c_2\alpha)+c_3,
\end{equation}
where $c_1,c_2$ and $c_3$ are the fitting coefficients tabulated in Table~\ref{table:resccoe_fit} for different intermolecular potentials.
As a consequence, the KLF at arbitrary TMAC can be roughly estimated by multiplying Eq.~\eqref{eq:endpoint_fit} and the rescaled KLF, with the maximum relative error being smaller than $10\%$. The KLF for the Cercignani-Lampis BC can also be rescaled according to the data in Table~\ref{table_cl_coe}.

%

 \begin{table}[t]
 	\caption{ Fitting coefficients of the velocity defect $U_s(0)$ at the solid surface by Eq.~\eqref{CL_fit} for inverse power-law potentials with different values of the viscosity index $\omega$,  when the diffuse-specular BC is used.  }
 	\label{table:resccoe_fit}
	\centering

	\begin{tabular}{ccccc}
\hline
	 & $\omega$
		&  $c_1$ & $c_2$  &   $c_3$   \\
\hline

		& 0.5
		& 1.798  & 0.2410  &  -1.1050   \\
		& 0.75
		& 1.776  & 0.2825  &   -1.0030   \\
		& 1
		& 1.750  & 0.3378  &  -0.8800   \\
		& 1.25
		&   1.820   & 0.3822  &    -0.8353   \\
			& 1.5
		&  1.895   & 0.4368  &  -0.7721   \\
\hline
	\end{tabular}\par
	
\end{table}

\section{Comparison with the experiment}\label{sec:exp}

\begin{figure}
	\begin{centering}
		\includegraphics[width=0.6\textwidth]{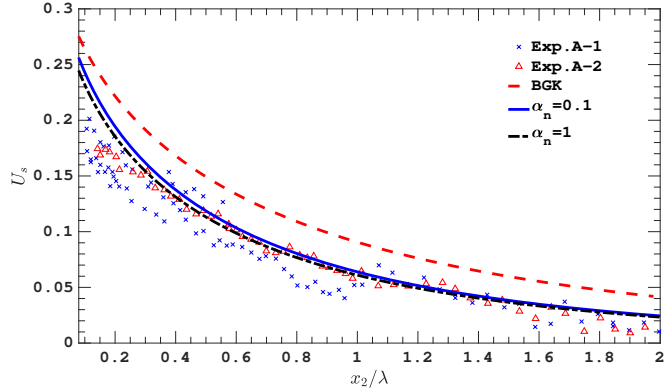}
		\par\end{centering}
	\caption{ Comparisons of the KLFs between the experiments and the LBE solution with $\alpha_t=0.88$, when the air molecules of $\omega=0.75$ and various $a_n$ are applied. Exp.~A1 and Exp.~A2 were measured by Reynolds et al.~\cite{reynolds1974velocity}. The diffuse-specular BC with $\alpha_M=0.88$ is used in the BGK model. }
	\label{fig:com_exp}
\end{figure}

Reynolds et al. measured the VSC and KLF for air passing along the surface of a highly polished aluminum plate~\cite{reynolds1974velocity} . They found that the KLF is different from the results predicted by the BGK model. Loyalka pointed out that such discrepancy is due to the deficiency of the BGK model~\cite{loyalka1975velocity}, where the collision frequency does not depend on the molecular velocity; by using a kinetic model with a variable collision frequency, a reasonable agreement of the velocity profile with the experimental data was observed. Given the apparent deficiency of the model equation, results from the LBE of HS molecules were also compared with the experimental data~\cite{ohwada1989numerical}.  However, all the previous works were based on the HS gas with an viscosity index of $\omega=0.5$, while air has an effective viscosity index of 0.75 at the room temperature. Moreover, the TMAC used in the numerical simulations was one, which results in a VSC of about one, while that measured by Reynolds et al. has an average value of $\bar{\zeta}_{Exp}=1.1$ (which has been corrected by multiplying a factor of $\sqrt{\pi}/2$)~\cite{reynolds1974velocity} .

In this section, we try to explain the experimental data using the LBE solutions for the inverse power-law potential with $\omega=0.75$. Although air is a mixture of oxygen and nitrogen, we treat it as a single-species monatomic gas, since (i) the molecular masses of oxygen and nitrogen are close to each other and (ii) for isothermal flow the mass flow rate (and hence the VSC) is insensitive to the rotational degrees of freedom~\cite{LeiJFM2015,Loyalka1979Polyatomic}.


Figure~\ref{fig:com_exp} shows the KLF obtained from the LBE with $\alpha_t=0.95$, and $\alpha_n=0.1$ and 1 under the Cercignani-Lampis BC, as well as the experimental data. The result from the BGK equation is also included for comparison. We use the value of TMAC $\alpha=0.95$, as our numerical calculation in the previous section suggests that the predicted VSC from the LBE agrees well with the experimental value of 1.1~\cite{reynolds1974velocity}. It is found that the KLF changes slightly under different $\alpha_n$, and the results of $\alpha_n=1$ seems better than the others in the agreement with the experimental data, while the solution of the BGK equation has a visible deviation from the experimental results. Note that when using the diffuse-specular BC, similar results can also be obtained for $\alpha_M=0.95$.

We note that the KLF from the experiments are scattered, which is inconsistent with the theoretical analysis that the normalized velocity near the solid surface should be independent of the mean free path and shear gradient. Reynolds et al. argued that the most possible reason was the inaccurate determination of the mean free path~\cite{reynolds1974velocity}. Therefore, intuitively, in order to interpret the experimental results, one should take this factor into account. To this end, we first assume the actual TMAC for the interaction of air with the polished aluminum plate is $\alpha_M$ in the diffuse-specular BC. Then we calculate the VSC $\bar{\zeta}(\alpha_M)$ from the LBE. If $\bar{\zeta}_{Exp}<\bar{\zeta}(\alpha_M)$, the mean free path in the experiment has been overestimated due to the inaccuracy in measuring the gas pressure. Therefore, the value of the KLF from the experimental should be multiplied by  $1/\sigma=\bar{\zeta}(\alpha_M)/\bar{\zeta}_{Exp}$, while the width of the KLF should be stretched by a factor of $1/\sigma$. In the numerical simulation, various values of $\alpha_M$ are attempted, until good agreement between the results of experiment and numerical simulation are achieved.

\begin{figure}\label{com_air}
	\begin{centering}
		\includegraphics[width=0.6\textwidth]{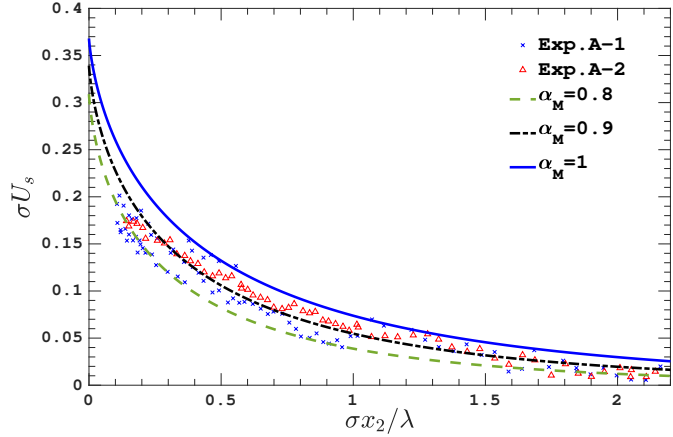}
		\par\end{centering}
	\caption{ Comparisons of the KLF between the experiments and the LBE solution with $\alpha_M=0.8,~0.9$ and 1, when the air molecules with $\omega=0.75$ are used. The LBE results are scaled by a factor $\sigma=\bar{\zeta}_{Exp}/\bar{\zeta}(\alpha_M)$, where $\bar{\zeta}_{Exp}=1.1$ is the average VSC from the experiments. Exp.~A1 and Exp.~A2 are measured by~\cite{reynolds1974velocity}. }
	\label{fig:com_exp_air}
\end{figure}

To show all the results in one figure, however, the KLF $U_s(x_2)$ obtained from the numerical simulation of the LBE has been rescaled to $\sigma{U_s}(\sigma{}x_2)$. Comparisons between the numerical and experimental results are depicted in Fig.~\ref{fig:com_exp_air}. It is seen that, when the TMAC varies from 0.8 to 1, the results of LBE can cover almost all the experimental data. In other words, the TMAC of the aluminum plate used in the air experiments is most likely $0.9\pm0.1$. If the TMAC is 0.9, we have $\sigma=0.9$, this means that the mean free path in the experiment is overestimated by $10\%$, which seems reasonable due to the accuracy of the micro-manometers at that time.

\section{Conclusions}\label{sec:summary}

In summary, we have proposed a synthetic iteration scheme to expedite the convergence of finding the steady-state solution of the linearized Boltzmann equation for the Couette flow between two parallel plates. 
In the free molecular flow regime, both the conventional and synthetic schemes lead to the same converged solution after several iterations. However, the synthetic iteration scheme converges significantly faster than the conventional one in the transition and near-continuum gas flow regimes, which is about two to three orders of magnitude faster than the conventional iterative scheme. Based on the Bhatnagar-Gross-Krook kinetic model, the synthetic iteration scheme is assessed to be accurate at least with six significant digits.

With this efficient and accurate method, the influences of the intermolecular potentials (i.e. the inverse power-law, Lennard-Jones, and shielded Coulomb potentials) and the kinetic boundary BCs on the Knudsen layer have been investigated based on the linearized Boltzmann equation, where the Boltzmann collision operator for general intermolecular potentials is solved by the fast spectral method. Both the diffuse-specular and  Cercignani-Lampis boundary conditions are considered.
It has been found that, although different intermolecular potentials lead to roughly the same value of the viscous slip coefficient, the KLF is strongly affected by the potential, whose value and width increase with the effective viscosity index of the gas.

The highly accurate VSC and its general relation to the TMAC are presented for different intermolecular potentials and gas-surface boundary conditions. In addition, the KLF is found to be perfectly fitted by the series $\sum_{n=0}^{2}\sum_{m=0}^{2}c_{n,m}x^n(x\ln{} x)^m$, where $x$ is the distance to the solid surface. Correspondingly, based on the obtained KLF, the macroscopic flow velocity gradient exhibits a logarithmic divergence on the boundary. The strength of this divergence depends on the coefficient $c_{0,1}$, whose value also increases with the viscosity index. Furthermore, the similarity of the KLF has been established by rescaling the KLF by the defect velocity at the solid surface. Consequently, the KLF at arbitrary TMAC can be predicted by multiplying the rescaled KLF and the defect velocity at the solid surface which is accurately fitted by an exponential function of the TMAC. These results are useful to formulate the effective shear viscosity~\cite{lockerby2005capturing} and slip boundary condition to be used in the framework of Navier-Stokes equations~\cite{lockerby2008modelling}.

The experimental data of the viscous slip coefficient and KLF measured by~\cite{reynolds1974velocity} has been interpreted fairly well by the linearized Boltzmann equation with a realistic viscosity index. We concluded that the TMAC for the interaction of air with the polished aluminum is most likely $0.9\pm0.1$, instead of $1.0$ as used in previous studies for a comparison with the experiment. This result suggests that mean free path in the experiment has been overestimated by about $10\%$.

Finally, it should be noted that the accurate and efficient synthetic iterative scheme developed in this paper are readily to be extended to multi-species gas mixtures~\cite{LeiJCP2017}. The influence of intermolecular potentials and gas kinetic boundary conditions on the Kramer's problem of gas mixtures are subject to future studies.

\section*{Acknowledgments}

This work is founded by joint project from the Royal Society of Edinburgh and National Natural Science Foundation of China under Grant No.~51711530130, the Carnegie Research Incentive Grant for the Universities in Scotland, and the Engineering and Physical Sciences Research Council (EPSRC) in the UK under grant EP/R041938/1.

\section*{References}
\bibliography{bibnew}
\end{document}